\documentclass[preprint,aps]{revtex4}
\usepackage{bm}
\usepackage{graphics}
\usepackage[dvips]{color}
\usepackage{dcolumn}
\usepackage{times}

\renewcommand{\r}{{\bf r}}
\newcommand{\R}{{\bf R}}

\newcommand{\rb}{{\bf r}_{\bot}}
\renewcommand{\k}{{\bf k}}
\newcommand{\K}{{\bf K}}

\renewcommand{\v}{{\bf v}}
\newcommand{\be}{\begin{eqnarray}}
\newcommand{\ee}{\end{eqnarray}}
\newcommand{\tn}{\tilde{n}}
\newcommand{\bn}{\bar{n}}

\newcommand{\tpsi}{\tilde{\psi}}
\newcommand{\tphi}{\tilde{\phi}}
\newcommand{\vphi}{\varphi}

\newcommand{\q}{{\bf q}}

\newcommand{\bu}{\bar{u}}
\newcommand{\bz}{\bar{z}}
\newcommand{\tu}{\tilde{u}}

\begin{document} 
\bibliographystyle{prsty}
\title{Coarse-Grained Finite-Temperature Theory for the Condensate in Optical Lattices }
\author{S. Konabe$^{1,2}$ and T. Nikuni$^{1}$}
\affiliation{$^{1}$Department of Physics, Faculty of Science, Tokyo University of Science, 1-3 Kagurazaka, Shinjuku-ku, Tokyo, 162-8601, Japan\\
$^{2}$Department of Physics, University of Toronto, Toronto, Ontario, M5S 1A7, Canada}

\begin{abstract}
In this work, we derive coarse-grained finite-temperature theory for a Bose condensate in a one-dimensional optical lattice, in addition to a confining harmonic trap potential. 
To construct the theory for the condensate and noncondensate in a periodic lattice potential, the difficulty arises due to the rapid variation in the position by a lattice potential, compared to the length scale of the harmonic potential. 
In order to overcome this difficulty, we need some coarse-graining procedure for the lattice potential. 
We start from a two-particle irreducible (2PI) effective action on the Schwinger-Keldysh closed-time contour path. 
In principle, this action involves all information of equilibrium and non-equilibrium properties of the condensate and noncondensate atoms. 
By assuming the {\it ansatz} for the variational function, i.e., the condensate order parameter in an effective action,
we derive a coarse-grained effective action, which describes the dynamics on the length scale much longer than a lattice constant. 
Using the variational principle, coarse-grained equations of motion for the condensate variables are obtained. These equations include a dissipative term due to collisions between condensate and noncondensate atoms, as well as noncondensate mean-field.
As a result of a coarse-graining procedure, the effects of a lattice potential are incorporated into equations of motion for the condensate by an effective mass, a renormalized coupling constant, and an umklapp scattering process. 
To illustrate the usefulness of our formalism, we discuss a Landau instability of the condensate in optical lattices by using the coarse-grained generalized Gross-Pitaevskii hydrodynamics. 
We found that the collisional damping rate due to collisions between the condensate and noncondensate atoms changes sign when the condensate velocity exceeds a renormalized sound velocity, leading to a Landau instability consistent with the Landau criterion.
Our results in this work give an insight into the microscopic origin of the Landau instability.
\end{abstract}

\maketitle    
\section{INTRODUCTION}
Recent extensive researches on ultracold atomic gases in optical lattices by experimental and theoretical approaches have revealed the nontrivial nature of many-body quantum systems,~\cite{bloch2005,morsch2006} such as the superfluid-Mott insulator transition,~\cite{greiner2002,xu2005} stability of superfluidity,~\cite{burger2001,cataliotti2003,fallani2004,fertig2005,sarlo2005} and the Josephson effect.~\cite{albiez2005} Because of the ease of  fine-tuning of experimental parameters, optical lattices are also used as testing grounds for many-body theory,~\cite{jaksch2005} including non-equilibrium phenomena,~\cite{altman2002,rey2004_2,rey2005_3,altman2005,kollath2007,temme2006} which are usually very difficult to analyze in traditional solid state systems.

Among a number of startling behaviors of the Bose condensate, superfluidity is one of the most fascinating phenomena. 
A long time ago, Landau showed that the superfluid state is stable as long as a velocity of a superfluid is smaller than a critical velocity, above which elementary excitations are spontaneously produced, making the superfluid state unstable.~\cite{landau1941,abrikosov1975} His argument relied only on the energy and momentum conservation and the Galilei transformation.
When a condensate is set in an optical lattice potential, the breakdown of superfluidity becomes more complicated due to the competition
between the inter-atomic interaction and the periodic lattice potential, as observed experimentally.
~\cite{burger2001,fallani2004,sarlo2005,ferlaino2002}
Theoretically, two different types of instability have been discussed within the Gross-Pitaevskii (GP) equation with a periodic lattice potential.~\cite{wu2001,wu2003,machholm2003,kramer2003,menotti2003,taylor2003,modugno2004,danshita2007}
One is  the Landau (or energetic) instability, which occurs when the excitation energy becomes negative. 
It is this instability that Landau originally argued.
The other is the dynamical instability, which occurs when the excitation energy possesses the imaginary part.
In this type of instability, the effect of the lattice potential is to couple an (unphysical) antiphonon to a phonon by 
the first order Bragg scattering, leading to the dynamical instability.~\cite{wu2001,wu2003,taylor2003}
It is important to note that the dynamical instability exists at zero temperature, and thus can be understood within the usual zero-temperature GP theory. 
In contrast, however, the Landau instability has been found to occur at finite temperatures.~\cite{sarlo2005}
As compared with the dynamical instability, the essential role of the lattice potential in the Landau instability is to pin the
incoherent thermally excited noncondensate, while the condensate can coherently tunnel through the lattice potential.~\cite{ferlaino2002} 
Thus the thermally excited atoms trapped by the optical lattice play a role of obstacles to the condensate, giving rise to dissipative effects.
For the above reason,  one cannot study microscopic mechanisms of the Landau instability by using the zero-temperature GP equation. One should include the effect of the thermal cloud into the theory.

In order to discuss the Landau instability in the periodic lattice, one needs a finite-temperature microscopic theory for the 
Bose-condensed gas including effects of the lattice potential. 
As far as very low temperature regime is concerned, the GP equation has succeeded in describing a 
trapped Bose-Einstein condensate.~\cite{dalfovo1999,pethic2002,pitaevskii2003}
At finite temperatures, however, the presence of the noncondensate in addition to the condensate makes the GP description insufficient. 
In order to incorporate the dynamics of the noncondensate,  a number of papers derived generalized GP equations at finite temperatures, which includes effects of noncondensates by mean-field and collisional exchange between condensate and noncondensate atoms, and a quantum kinetic equation for the noncondensate.~\cite{griffin1996,gardiner1998,walser1999,zaremba1999,stoof1999}
A resulting two-component many-body system leads to non-trivial phenomena such as nucleation
and evaporation of condensates,~\cite{walser1999,zaremba1999,stoof1999,stoof1997,proukakis1998} and
damping of collective modes,~\cite{shi1998,pitaevskii1997,fedichev1998,giorgini1998} which do not appear in the GP theory. 
It is thus natural to anticipate that incorporation of the periodic lattice potential into the coupled many-body system of the condensate
and noncondensate atoms will lead to interesting new physics.
The main purpose of the present paper is to show one of the crucial effects due to thermally excited noncondensate atoms in 
optical lattices by focusing on the breakdown of superfluidity.

In the present paper, for investigating non-equilibrium dynamics of  such a two-component many-body system in the
periodic optical lattice potential, we construct a finite-temperature theory for the condensate in the one-dimensional
optical lattice, in addition to a confining three-dimensinal trap potential.
For this purpose, we start from the 2PI effective action~\cite{cornwall1974} with the Schwinger-Keldysh closed-time path formalism.
~\cite{schwinger1961,keldysh1964,danielewicz1984}
There are two advantages to use a functional integral formalism.
The one is that one can derive an action in the non-equilibrium quantum
field theory by controllable approximations.~\cite{rey2004_2,rey2005_3,berges2004,gasenzer2005}
The other is that one can introduce an {\it ansatz} as the variational function, i.e., the order parameter of the condensate in the effective action and perform coarse-grained approximation in a well-defined manner since the action involves integrations over position. 
Making coarse-grained procedure, one can include the effects of the lattice potential effectively.
By means of the coarse-grained effective action, we obtain coarse-grained equations of motion for the condensate variables at finite temperatures including
the effect of the optical lattice. 
As an application of our finite-temperature coarse-grained theory, we investigate the 
microscopic mechanism of the breakdown of the superfluidity having experiments such as reported in Ref.~\onlinecite{sarlo2005} in mind.

Recent papers~\onlinecite{konabe2006_2,iigaya2006,konabe2007_1}  reported the earlier attempts to study the breakdown of superfluidity in an optical lattice, focusing on microscopic mechanisms. 
When the condensate has finite velocity, the damping rate of the collective oscillation, which stems from the collisional or Landau damping processes, can change its sign at a critical velocity. 
This means that the inverse process of damping occurs at the critical velocity, resulting in spontaneous production of elementary excitations. 
Moreover, the increase of excitations in time as a result of inverse damping process suggests the breakdown of the stable superfluid state.
In Refs.~\onlinecite{konabe2006_2,iigaya2006,konabe2007_1}, starting from 
the one-dimentional Bose-Habbard model, the authors derived equations of motion for the
condensate order parameter at finite temperatures, which include the effect of the 
noncondensate atoms through  mean-field interactions as well as collisions.
From these equations of motion, the authors calculated damping rate of collective modes
(phonons) in the collisionless regime of interest and showed 
that the damping rate changes its sign at a critical velocity.
This instability is shown to coincide with the Landau instability.

In contrast to the previous works,~\cite{konabe2006_2,iigaya2006,konabe2007_1}
in the present paper we use a coarse-grained theory.
It will be shown that the coarse-grained formalism developed here describes the breakdown
mechanism in a more transparent way.
From the condition of the negative damping rate, we automatically obtain both the
negative excitation energy and the Landau criterion, which is modified by the lattice potential. 

The present paper is organized as follows. 
In Sec.\ref{sec:2PI}, we briefly review the non-equilibrium quantum field theory that consists of the 2PI effective action with the Schwinger-Keldysh closed-time formalism. In Sec.~\ref{sec:review}, for illustration of the approximation, we derive a generalized GP equation for the condensate, which can be written as the hydrodynamic equations in terms of the local condensate density and the superfluid velocity. In Sec.~\ref{sec:generalized_hydrodynamic}, we derive the coarse-grained 
effective action for the Bose gas in optical lattices in terms of coarse-grained macroscopic variables. 
Using the variational principle, we obtain coarse-grained equations of motion of the condensate variables suitable for describing the condensate at finite temperatures in an optical lattice. In Sec.~\ref{sec:instability}, in order to show the usefulness of our theory, we apply coarse-grained generalized GP hydrodynamic equations to discuss the breakdown of the superfluidity and give the microscopic origin of the Landau instability.
 
\section{2PI EFFECTIVE ACTION AND NON-EQUILIBRIUM QUANTUM FIELD THEORY}\label{sec:2PI}
We consider a Bose-condensed gas trapped in a one-dimensional optical lattice
in addition to the harmonic potential: 
\begin{eqnarray}
&&V_{\rm ext}({\bf r})=V_{\rm ho}({\bf r})+V_{\rm opt}(z),\\
&&V_{\rm ho}({\bf r})=\frac{m}{2}(\omega_x^2x^2+\omega_y^2y^2+\omega_z^2z^2),\\
&&V_{\rm opt}(z)=sE_{R}\cos^2\left(\frac{\pi}{d}z\right),\label{2PI:potential}
\end{eqnarray}
where $\omega_x$, $\omega_y$, $\omega_z$ are frequencies of the harmonic trap potential, $s$ is the dimensionless parameter describing strength of the lattice potential, $E_{R}=\hbar^2/2md^2$ is the recoil energy,
 and $d$ is the lattice constant.
A Bose gas in the external potential represented by Eq.~(\ref{2PI:potential}) is described by the following Lagrangian density
\begin{eqnarray}
\mathcal{L}({\bf r},t)&=&\psi^*({\bf r},t)i\hbar\frac{\partial}{\partial t}\psi({\bf r},t)+\psi^*({\bf r},t)\frac{\hbar^2\nabla^2}{2m}\psi({\bf r},t)\nonumber\\
&&{}-V_{{\rm ext}}({\bf r})|\psi({\bf r},t)|^2
-\frac{g}{2}|\psi({\bf r},t)|^4,
\label{2PI:Lagrangian}
\end{eqnarray}
where $\psi$ is the scalar field describing Bose atoms. 
We have assumed that the interaction between atoms is represented as a short-range peudopotential with the coupling constant $g$, which is related to the
$s$-wave scattering length of atoms through $g=4\pi \hbar^2 a/m$. 


\subsection{Generating Functional in the Non-equilibrium Quantum Field Theory}\label{subsec:generating_functional}
An efficient way to treat non-equilibrium dynamics~\cite{berges2004} is using the two-particle irreducible (2PI) effective action~\cite{cornwall1974} with the Schwinger-Keldysh
closed-time-path formalism.~\cite{schwinger1961,keldysh1964,danielewicz1984}
This formalism provides a powerful starting point for systematic approximations in the non-equilibrium quantum field theory.~\cite{rey2004_2,berges2004,gasenzer2005} For the ultracold atomic Bose gases, Rey {\it et al.} first applied this method to study various approximations and to develop the quantum kinetic theory for the condensate in optical lattices.~\cite{rey2004_2,rey2005_3}
From the 2PI effective action derived below, one can obtain a generalized GP equation for the condensate order parameter including effects of noncondensate atoms as a dissipative term and the noncondensate mean-field, and some kind of a kinetic equation for noncondensate atoms on an equal footing. 
In this section,
we briefly review this technique (see Ref.~\onlinecite{berges2004} for more details). We use units where $\hbar=1$ in this section.

\begin{figure}
 \begin{center}
      \scalebox{0.2}[0.2]{\includegraphics{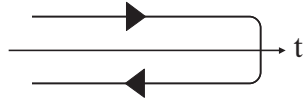}} 
    \caption{The Schwinger-Keldysh contour path}
    \label{fig:Keldysh_path}
  \end{center}
\end{figure}

In the quantum field theory, any correlation functions, which contain all information about a non-equilibrium many-body system, are obtained from a generating functional. As long as the initial density matrix is approximated by the Gaussian form, the generating functional in the non-equilibrium field theory can be written as a functional integral~\cite{berges2004,gasenzer2005}
\begin{eqnarray}
Z[{\bm J},{\bm K}]
=\int\mathcal{D}{\bm \psi}\exp\left[i\left(S[\hat{\psi}]+{\bm J}^{\dag}{\bm \psi}+\frac{1}{2}{\bm \psi}^{\dag}\hbar{\bm K}{\bm \psi}\right)\right],\label{2PI:generating_functional}
\end{eqnarray} 
where we have introduced matrix notation for the scalar field ${\bm \psi}$ and the source field ${\bm J}$
\begin{eqnarray}
{\bm \psi}({\bf r},t)&\equiv&(\psi({\bf r},t), \psi^*({\bf r},t))^t\equiv (\psi_1({\bf r},t),\psi_2({\bf r},t))^t,\nonumber\\
{\bm J}({\bf r},t)&\equiv&(J({\bf r},t), J^*({\bf r},t))^t\equiv (J_1({\bf r},t), J_2({\bf r},t))^t,
\end{eqnarray}
and suppressed the integration over space and time variables. 
A $2\times 2$ matrix nonlocal source field ${\bm K}$ is added to the action in order to obtain equations of motion for two-point correlation functions (Green's functions). 
The classical action is defined by the Lagrangian density (\ref{2PI:Lagrangian})
\begin{eqnarray}
S[{\bm \psi}]=\int d{\bf r}\int_{\mathcal{C}} dt\ \mathcal{L}({\bf r},t),\label{2PI:classical_action}
\end{eqnarray}
where the subscript $\mathcal{C}$ of the time-integration means that the integral is performed on the Schwinger-Keldysh contour path, which extends from the initial time $t_0$ to the finite time $t>t_0$, and back from $t$ to $t_0$ (Fig.\ref{fig:Keldysh_path}). 
From the generating functional~(\ref{2PI:generating_functional}), one can define a more useful generating functional for the connected Green's function
\begin{eqnarray}
W[{\bm J},{\bm K}]=-i\ln Z[{\bm J},{\bm K}].\label{2PI:connected_generating_functional}
\end{eqnarray}
The derivative of the generating functional $W$ with respect to the source field ${\bm J}$ gives the condensate order parameter
\begin{eqnarray}
\frac{\delta W[{\bm J},{\bm K}]}{\delta J_i({\bf r},t)}=\Phi_i({\bf r},t)\qquad (i=1,2).\label{2PI:derivative_W_J}
\end{eqnarray}
This order parameter $\Phi({\bf r},t)$ is the average of the original complex field ${\bm \psi}$ taken by the action~(\ref{2PI:classical_action}) :
\begin{eqnarray}
{\bm \Phi}({\bf r},t)&\equiv&(\Phi({\bf r},t),\Phi^*({\bf r},t))^{t}\nonumber\\
&\equiv&(\Phi_1({\bf r},t),\Phi_2({\bf r},t))^{t}\nonumber\\
&\equiv&\int\mathcal{D}{\bm \psi}\ {\bm \psi}\exp\left[i\left(S[{\bm \psi}]+{\bm J}^{\dag}{\bm \psi}+\frac{1}{2}{\bm \psi}^{\dag}\hbar{\bm K}{\bm \psi}\right)\right]\nonumber\\
&\equiv&\langle{\bm \psi}({\bf r},t)\rangle.
\end{eqnarray}
On the other hand, the derivative of $W[{\bm J},{\bm K}]$ with respect to the nonlocal source field ${\bm K}$ gives
\begin{eqnarray}
\frac{\delta W[{\bm J},{\bm K}]}{\delta {\bm K}({\bf r},t;{\bf r}',t')}=\frac{1}{2}\left[{\bm H}({\bf r},t;{\bf r}',t')+i{\bm G}({\bf r},t;{\bf r}',t')\right].\label{2PI:derivative_of_W_K}
\end{eqnarray}
Here, we have defined the condensate matrix Green's function, ${\bm H}({\bf r},t;{\bf r}',t')$, and the noncondensate connected matrix Green's function, ${\bm G}({\bf r},t;{\bf r}',t')$:
\begin{eqnarray}
{\bm H}({\bf r},t;{\bf r}',t')=\left[
\begin{array}{cc}
\Phi({\bf r},t)\Phi^*({\bf r}',t') & \Phi({\bf r},t)\Phi({\bf r}',t')\\
\Phi^*({\bf r},t)\Phi^*({\bf r}',t') & \Phi^*({\bf r},t)\Phi({\bf r}',t')
\end{array}
\right],
\end{eqnarray}
\begin{eqnarray}
i{\bm G}({\bf r},t;{\bf r}',t')=\left[
\begin{array}{cc}
\langle \tpsi({\bf r},t)\tpsi^*({\bf r}',t')\rangle & \langle \tpsi({\bf r},t)\tpsi({\bf r}',t')\rangle \\
\langle \tpsi^*({\bf r},t)\tpsi^*({\bf r}',t')\rangle & \langle \tpsi^*({\bf r},t)\tpsi({\bf r}',t')\rangle
\end{array}
\right].\label{2PI:matrix_Green's_function}
\end{eqnarray}
In the noncondensate Green's function, giving by (\ref{2PI:matrix_Green's_function}), we introduced the noncondensate field:
\be
\tilde{\bm \psi}(\r,t)&\equiv&(\tilde\psi(\r,t),\tilde\psi^*(\r,t))^{t}\nonumber\\
&\equiv&(\psi(\r,t)-\Phi(\r,t),\psi^*(\r,t)-\Phi^*(\r,t))^{t}
\ee
We note that averages over the fields are automatically time ordered on the closed-time path in the functional integral.

We now define the effective action, which is the generating functional for the two-particle irreducible vertex functions, through the Legendre transform
\begin{eqnarray}
\Gamma[{\bm \Phi},{\bm G}]=W[{\bm J},{\bm K}]-{\bm J}^{\dag}{\bm \Phi}-\frac{1}{2}{\bm \Phi}^{\dag}{\bm K}{\bm \Phi}-\frac{i}{2}{\rm Tr}\left[{\bm G} {\bm K}\right].\label{2PI:effective_action_Legendre}
\end{eqnarray}
Following Ref.~\onlinecite{cornwall1974}, one obtains the expression for the effective 
action $\Gamma[{\bm \Phi},{\bm G}]$ as
\begin{eqnarray}
\Gamma[{\bm \Phi},{\bm G}]&=&S[{\bm \Phi}]+\frac{i}{2}{\rm Tr}\ln{\bm G}^{-1}+\frac{i}{2}{\rm Tr}\left[{\bm D}^{-1}{\bm G}\right]+\Gamma_2[{\bm \Phi},{\bm G}]+{\rm Const.},\nonumber\\
\label{2PI:2PI_effective_action}
\end{eqnarray}
where ${\bm D}^{-1}$ is the classical inverse propagator matrix defined by
\begin{eqnarray}
{\bm D}^{-1}({\bf r},t;{\bf r}',t')\equiv\frac{\delta^2 S[{\bm \Phi}]}{\delta{\bm \Phi}({\bf r},t)\delta{\bm \Phi}^{\dag}({\bf r}',t')},\label{2PI:classical_propagator}
\end{eqnarray}
with the $2\times 2$ matrix elements
\begin{eqnarray}
D_{11}^{-1}({\bf r},t;{\bf r}',t')&=&\frac{\delta^2 S[{\bm\Phi}]}{\delta\Phi^*({\bf r},t)\delta\Phi({\bf r}',t')}\nonumber\\
&=&\left[i\frac{\partial}{\partial t}-\mathcal{H}_0({\bf r})\right]\delta({\bf r}-{\bf r}')\delta(t-t')\nonumber\\
&&{}-2g|\Phi({\bf r})|^2\delta({\bf r}-{\bf r}'),\label{2PI:classical_propagator_11}\\
D_{12}^{-1}({\bf r},t;{\bf r}',t')
&=&\frac{\delta^2 S[{\bm\Phi}]}{\delta\Phi({\bf r},t)\delta\Phi({\bf r}',t')}\nonumber\\
&=&-2g\left[\Phi({\bf r})\Phi({\bf r}')\right]^2\delta({\bf r}-{\bf r}'),\label{2PI:classical_propagator_12}\\
D_{21}^{-1}({\bf r},t;{\bf r}',t')
&=&\frac{\delta^2 S[{\bm\Phi}]}{\delta\Phi^*({\bf r},t)\delta\Phi^*({\bf r}',t')}\nonumber\\
&=&-2g\left[\Phi^*({\bf r})\Phi^*({\bf r}')\right]^2\delta({\bf r}-{\bf r}'),\label{2PI:classical_propagator_21}\\
D_{22}^{-1}({\bf r},t;{\bf r}',t')&=&\frac{\delta^2 S[{\bm\Phi}]}{\delta\Phi({\bf r},t)\delta\Phi^*({\bf r}',t')}\nonumber\\
&=&\left[-i\frac{\partial}{\partial t}-\mathcal{H}_0({\bf r})\right]\delta({\bf r}-{\bf r}')\delta(t-t')\nonumber\\
&&{}-2g|\Phi({\bf r})|^2\delta({\bf r}-{\bf r}').\label{2PI:classical_propagator_22}
\end{eqnarray}
In Eqs.~(\ref{2PI:classical_propagator_11})$-$(\ref{2PI:classical_propagator_22}), $\mathcal{H}_0({\bf r})$ is defined by a one-body part
\begin{eqnarray}
\mathcal{H}_0({\bf r})\equiv-\frac{1}{2m}\nabla^2+V_{{\rm ext}}({\bf r}).
\end{eqnarray}
The trace and logarithm in Eq.~(\ref{2PI:2PI_effective_action}) is defined by the functional integral.
$\Gamma_2[{\bm \Phi},{\bm G}]$ in Eq.~(\ref{2PI:2PI_effective_action}) consists of two-particle irreducible vacuum diagrams (the diagrams that cannot be disconnected by cutting two propagator lines) with full propagators ${\bm G}$. The vertices are determined by the interaction term $S_{\rm int}[{\bm \Phi},\tilde{\bm \psi}]$, which is the part higher than second order in $\psi$ of the expansion for the action $S[{\bm \Phi}+\tilde{\bm \psi}]$.

In this paper, we consider a relatively high-temperature regime by treating noncondensate atoms within the Hartree-Fock approximation, neglecting off-diagonal components of the
Green's functions. The resulting effective action is given by
\begin{eqnarray}
\Gamma[{\bm \Phi},{\bm G}]=\Gamma_{{\bm \Phi}}[{\bm \Phi},{\bm G}]+\Gamma_{{\bm G}}[{\bm \Phi},{\bm G}],\label{2PI:HF_action}
\end{eqnarray}
where
\begin{eqnarray}
\Gamma_{{\bm \Phi}}[{\bm \Phi},{\bm G}]
&\equiv&S_{\rm GGP}[{\bm \Phi},{\bm G}]\nonumber\\
&=&\int d{\bf r}\int_c dt  \Phi^*({\bf r},t)\biggl\{
i\frac{\partial}{\partial t}-\mathcal{H}_0({\bf r})-\frac{g}{2}\left |\Phi({\bf r},t)\right |^2\nonumber\\
&&{}-ig\biggl[G_{11}({\bf r},t;{\bf r},t)+G_{22}({\bf r},t;{\bf r},t)\biggl]\biggl\}\Phi({\bf r},t)\nonumber\\
&&{}+\Gamma_2[{\bm \Phi},{\bm G}],\label{2PI:Gamma_phi}\\
\Gamma_{{\bm G}}[{\bm \Phi},{\bm G}]&=&\frac{1}{2}\int d{\bf r} \int_c dt\left[i\frac{\partial}{\partial t}-\mathcal{H}_0({\bf r})\right]i G_{11}({\bf r},t;{\bf r},t)\nonumber\\
&&+\frac{1}{2}\int d{\bf r} \int_c dt\left[-i\frac{\partial}{\partial t}-\mathcal{H}_0({\bf r})\right]i G_{22}({\bf r},t;{\bf r},t)\nonumber\\
&&+\frac{i}{2}{\rm Tr}\ln {\bm G}^{-1}.\label{2PI:Gamma_G}
\end{eqnarray}
In Eq.~(\ref{2PI:HF_action}), the matrix Green's function contains only diagonal components because we will use the Hartree-Fock approximation for the noncondensate. The subscript ${\rm GGP}$ of $S_{\rm GGP}$ in Eq.~(\ref{2PI:Gamma_phi}) indicates that $S_{\rm GGP}$ will be shown to lead to the generalized GP equation.

Taking the derivative of the effective action~(\ref{2PI:HF_action}) with respect to ${\bm \Phi}$ and ${\bm G}$, one obtains $\delta \Gamma/\delta {\bm \Phi}=-{\bm J}-{\bm K}{\bm \Phi}$ and $\delta \Gamma/\delta {\bm G}=-i{\bm K}/2$, respectively.
In a real physical state, the artificial external fields ${\bm J}$ and ${\bm K}$ should vanish. 
This requirement yields equations of motion
$\delta \Gamma/\delta \Phi^*({\bf r},t)=\delta S_{\rm GGP}/\delta\Phi^*(\r,t)=0$:
\begin{eqnarray}
&&i\frac{\partial}{\partial t}\Phi({\bf r},t)=\mathcal{H}_0({\bf r})\Phi({\bf r},t)+g\biggl[|\Phi({\bf r},t)|^2+iG_{11}({\bf r},t;{\bf r},t)\nonumber\\
&&{}\qquad\qquad\qquad+iG_{22}({\bf r},t;{\bf r},t)\biggl]\Phi({\bf r},t)
-\frac{\delta \Gamma_2[{\bm \Phi},{\bm G}]}{\delta \Phi^*({\bf r},t)},\label{2PI:generalized_GP}
\ee
and 
$\delta \Gamma/\delta G_{11}({\bf r},t;{\bf r}',t')=0$:
\be
G_{11}^{-1}({\bf r},t;{\bf r}',t')=D_{11}^{-1}({\bf r},t;{\bf r}',t')-\Sigma_{11}({\bf r},t;{\bf r}',t').\label{2PI:Dyson_Beliaev}
\end{eqnarray}
Eq.~(\ref{2PI:generalized_GP}) is the equation of motion for the condensate order parameter, while Eq.~(\ref{2PI:Dyson_Beliaev}) is the non-equilibrium Dyson equation for the noncondensate atoms with the proper self-energy defined by
\begin{eqnarray}
\Sigma_{11}({\bf r},t;{\bf r}',t')\equiv 2i\frac{\delta\Gamma_2[{\bm \Phi},{\bm G}]}{\delta G_{11}({\bf r},t;{\bf r}',t')}.\label{2PI:self_energy}
\end{eqnarray}
Note that since the time integration in Eq.~(\ref{2PI:Dyson_Beliaev}) is defined on the Schwinger-Keldysh contour path, this equation is generalization of the usual Dyson equation to the non-equilibrium systems.
After projecting the time-integration on the Schwinger-Keldysh contour path onto the real-time contour, Eq.~(\ref{2PI:Dyson_Beliaev}) leads to a kinetic equation for the noncondensate distribution function.~\cite{haug1996}

\begin{figure}
  \begin{center}
      \scalebox{0.5}[0.5]{\includegraphics{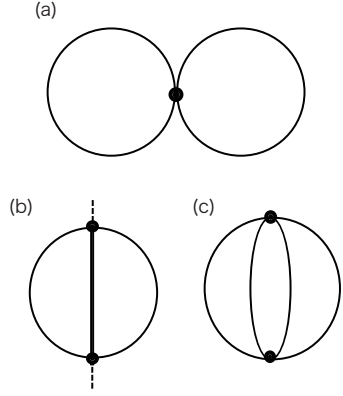}} 
    \caption{2PI diagrams up to second order in the coupling constant; the double-bubble (a), the setting sun (b), and the basketball (c).}
    \label{fig:diagram}
  \end{center}
\end{figure}

In order to obtain the 2PI part $\Gamma_2$, one has to perform approximations suitable for the physical problem under consideration by truncating diagrammatic expansions.
We show diagrams for $\Gamma_2$ used in the present analysis in Fig.~\ref{fig:diagram}, where two- and three-loop vacuum diagrams are illustrated. The vertices are specified by the interaction $S_{{\rm int}}[{\bm \Phi},\tilde{\bm \psi}]$. 
Analytical expressions corresponding to Fig.~\ref{fig:diagram} are
\begin{eqnarray}
\Gamma_2[{\bm \Phi},{\bm G}]=\Gamma_2^{(1)}[{\bm \Phi},{\bm G}]
+\Gamma_2^{(2:{\rm c})}[{\bm \Phi},{\bm G}]
+\Gamma_2^{(2:{\rm nc})}[{\bm \Phi},{\bm G}],\label{2PI:2PI_action}
\end{eqnarray}
where
\begin{eqnarray}
\Gamma_2^{(1)}[{\bm \Phi},{\bm G}]&=&\frac{g}{8}\int d{\bf r}\int_c dt
\biggl[
G_{11}({\bf r},t;{\bf r},t)G_{11}({\bf r},t;{\bf r},t)\nonumber\\
&&{}\quad\ +6G_{11}({\bf r},t;{\bf r},t)G_{22}({\bf r},t;{\bf r},t)\nonumber\\
&&{}\quad\ +G_{22}({\bf r},t;{\bf r},t)G_{22}({\bf r},t;{\bf r},t)\biggl],\label{2PI:Gamma_2_a}\\
\Gamma_2^{(2:{\rm c})}[{\bm \Phi},{\bm G}]&=&-\frac{g^2}{4}\int d{\bf r} d{\bf r}'\int_c dtdt'
\biggl[H_{11}({\bf r},t;{\bf r}',t')G_{11}({\bf r},t;{\bf r}',t')\nonumber\\
&&{}\quad\times G_{22}({\bf r},t;{\bf r}',t')G_{22}({\bf r},t;{\bf r}',t')\nonumber\\
&&{}\quad+H_{22}({\bf r},t;{\bf r}',t')G_{22}({\bf r},t;{\bf r}',t')\nonumber\\
&&{}\quad\times G_{11}({\bf r},t;{\bf r}',t')G_{11}({\bf r},t;{\bf r}',t')\biggl],\label{2PI:Gamma_2_b}\\
\Gamma_2^{(2:{\rm nc})}[{\bm \Phi},{\bm G}]&=&-\frac{g^2}{8}\int d{\bf r} d{\bf r}'\int_c dtdt'
\biggl[G_{11}({\bf r},t;{\bf r}',t')G_{11}({\bf r},t;{\bf r}',t')\nonumber\\
&&{}\quad\times G_{22}({\bf r},t;{\bf r}',t')G_{22}({\bf r},t;{\bf r}',t')\biggl].\label{2PI:Gamma_3}
\end{eqnarray}
From these equations, self-energies are derived by using the relation~(\ref{2PI:self_energy}):
\begin{eqnarray}
\Sigma_{11}^{(1)}({\bf r},t;{\bf r}',t')
&=&\frac{ig}{2}\biggl[G_{11}({\bf r},t;{\bf r}',t')+G_{22}({\bf r},t;{\bf r}',t')\biggl]\delta(t-t')\delta({\bf r}-{\bf r}')\nonumber\\
&&{}+ig G_{22}({\bf r},t;{\bf r}',t')\delta(t-t')\delta({\bf r}-{\bf r}'),\label{2PI:self-energy_1st}\\
\Sigma_{11}^{(2:{\rm c})}({\bf r},t;{\bf r}',t')
&=&-\frac{ig^2}{2}\biggl[H_{11}{({\bf r}',t';{\bf r},t)}G_{11}({\bf r},t;{\bf r}',t')G_{11}({\bf r},t;{\bf r}',t')\nonumber\\
&&{}+2H_{11}({\bf r},t;{\bf r}',t')G_{11}({\bf r},t;{\bf r}',t')G_{11}({\bf r}',t';{\bf r},t)\biggl],\label{2PI:self-energy_2nd_c}\\
\Sigma_{11}^{(2:{\rm nc})}({\bf r},t;{\bf r}',t')
&=&-2ig^2G_{11}({\bf r},t;{\bf r}',t')G_{11}({\bf r},t;{\bf r}',t')G_{11}({\bf r}',t';{\bf r},t).\label{2PI:self-energy_2nd_nc}
\end{eqnarray}

Before closing this section, we mention some approximations for the effective action.~\cite{rey2004_2,rey2005_3,berges2004,gasenzer2005}
The GP equation is obtained by retaining only the classical action $S[{\bm \Phi}]$ in
Eq.~(\ref{2PI:2PI_effective_action}).
This corresponds to the mean-field approximation appropriate at zero temperature.
When one retains all terms except $\Gamma_2$, this yields the Bogoliubov or one-loop approximation. Including $\Gamma_2$ up to first order in the coupling constant $g$, one obtains the time-dependent Hartree-Fock-Bogoliubov equations.~\cite{griffin1996}
One needs to consider higher-order approximation to include the multiple scattering effect.~\cite{rey2004_2,berges2004,gasenzer2005}
It will be shown that the multiple scattering gives rise to the dissipative term in the
equation of motion for the condensate and the colllision integrals in the kinetic equation.

\section{MICROSCOPIC FINITE-TEMPERATURE THEORY FOR THE BOSE GASES: A REVIEW}\label{sec:review}
It is instructive to derive the generalized GP equation and hydrodynamic equations for the condensate at finite temperatures, which have been derived in Ref.~\onlinecite{zaremba1999}, within the 2PI formalism.
In this section, we give a derivation of a generalized GP equation by specifying diagrams to approximate $\Gamma_2$.
In Sec.~\ref{sec:generalized_hydrodynamic}, we will use the technique discussed in this section to derive generalized hydrodynamic equations including the effect of optical lattices.

\subsection{Generalized Gross-Pitaevskii Equation}\label{subsec:generalized_GP}
Using Eqs.~(\ref{2PI:Gamma_2_a}), (\ref{2PI:Gamma_2_b}), and (\ref{2PI:Gamma_3}),
one obtains the analytical expression of the last term of Eq.~(\ref{2PI:generalized_GP}) as
\begin{eqnarray}
\frac{\delta\Gamma_{2}[{\bm \Phi},{\bm G}]}{\delta\Phi^*({\bf r},t)}&=&-\frac{g^2}{2}\int d{\bf r}'\int_c dt'\nonumber\\
&&{}\times G_{11}({\bf r},t;{\bf r}',t')G_{11}({\bf r},t;{\bf r}',t')G_{11}({\bf r}',t';{\bf r},t)\Phi({\bf r}',t')\nonumber\\
&\equiv& -\int d{\bf r}'\int_c dt' F({\bf r},t;{\bf r}',t')\Phi({\bf r}',t'),\label{generalized_GP:derivative_of_Gamma2_HF}
\end{eqnarray}
where we have used the relation $G_{22}({\bf r},t;{\bf r}',t')=G_{11}({\bf r}',t';{\bf r},t)$. 
For later convenience, we have introduced the following quantity 
\begin{eqnarray}
F({\bf r},t;{\bf r}',t')\equiv\frac{g^2}{2}G({\bf r},t;{\bf r}',t')G({\bf r},t;{\bf r}',t')G({\bf r}',t';{\bf r},t).\label{generalized_GP:F}
\end{eqnarray}
Here and hereafter, we use the notation $G({\bf r},t;{\bf r}',t')$ instead of $G_{11}({\bf r},t;{\bf r}',t')$.
In Eq.~(\ref{generalized_GP:derivative_of_Gamma2_HF}), it should be noted that the time integration is defined on the Schwinger-Keldysh contour path. In order to perform the integration in Eq.~(\ref{generalized_GP:derivative_of_Gamma2_HF}) explicitly, one has to project the time integration on the Schwinger-Keldysh contour path onto the real time axis. 
By virtue of the principle of causality, the integrand in Eq.~(\ref{generalized_GP:derivative_of_Gamma2_HF}) is replaced with the retarded counterpart. After using the Langreth theorem,~\cite{haug1996} one obtains
\begin{eqnarray}
&&-\int_c dt' F({\bf r},t;{\bf r}',t')\Phi({\bf r}',t')\nonumber\\
&&{}=
-\frac{g^2}{2}\int_{-\infty}^{\infty} dt'\biggl[
G^{(+)}({\bf r},t;{\bf r}',t')G^<({\bf r},t;{\bf r}',t')G^<({\bf r}',t';{\bf r},t)\nonumber\\
&&{}\qquad\qquad\qquad+G^<({\bf r},t;{\bf r}',t')G^{(+)}({\bf r},t;{\bf r}',t')G^<({\bf r}',t';{\bf r},t)\nonumber\\
&&{}\qquad\qquad\qquad+G^{(+)}({\bf r},t;{\bf r}',t')G^{(+)}({\bf r},t;{\bf r}',t')G^<({\bf r}',t';{\bf r},t)\nonumber\\
&&{}\qquad\qquad\qquad+G^<({\bf r},t;{\bf r}',t')G^<({\bf r},t;{\bf r}',t')G^{(-)}({\bf r}',t';{\bf r},t)\biggl]\nonumber\\
&&{}\qquad\qquad\qquad\times\Phi({\bf r}',t'),
\label{generalized_GP:causality}
\end{eqnarray}
where we have introduced the lesser, greater, retarded, and advanced Green's functions as
\be
&&G^{<}(\r,t;\r',t')\equiv-i\langle \tpsi^{*}(\r',t')\tpsi(\r,t)\rangle,\\
&&G^{>}(\r,t;\r',t')\equiv-i\langle \tpsi(\r,t)\tpsi^{*}(\r',t')\rangle,\\
&&G^{(+)}(\r,t;\r,t')\equiv-i\theta(t-t')\langle [\tpsi(\r,t),\tpsi^{*}(\r',t')]\rangle,\\
&&G^{(-)}(\r,t;\r',t')\equiv i\theta(t'-t)\langle[\tpsi(\r,t),\tpsi^{\*}(\r',t')]\rangle.
\ee

Eq.~(\ref{generalized_GP:causality}) involves terms that are nonlocal in space and time, which make it difficult to solve the equation. 
As in Refs.~\onlinecite{zaremba1999,stoof1999}, we assume that the macroscopic variables vary slowly in space and time. We thus approximate the condensate order parameter near the specific position and time $(\r,t)$ as 
\begin{eqnarray}
\Phi({\bf r}',t')&=&\sqrt{n_c({\bf r}',t')}e^{i\theta({\bf r}',t')}\nonumber\\
&\simeq&\sqrt{n_c({\bf r},t)}e^{i\left[\theta({\bf r},t)+\partial_t\theta({\bf r},t)(t'-t)+\nabla\theta({\bf r},t)\cdot({\bf r}'-{\bf r})\right]}\nonumber\\
&\equiv&\Phi({\bf r},t)e^{-i\left[\omega_c({\bf r},t)(t'-t)-{\bf k}_c({\bf r},t)\cdot({\bf r}'-{\bf r})\right]},\label{2PI:markovian}
\end{eqnarray}
where $n_c({\bf r},t)$ and $\theta({\bf r},t)$ are the condensate density and the phase of the order parameter, respectively. The condensate frequency and wavevector are defined by $\omega_c({\bf r},t)=-\partial_t\theta({\bf r},t)$ and ${\bf k}_c({\bf r},t)=\nabla\theta({\bf r},t)$, respectively. 
Next, we perform the gradient expansion for the noncondensate Green's functions in Eq.(\ref{generalized_GP:causality}) in order to separate the scale.
For this purpose, we introduce the relative coordinate and time and the center of mass coordinate and time
\begin{eqnarray}
&&\bar{{\bf r}}\equiv {\bf r}-{\bf r}',\qquad {\bf R}\equiv \frac{{\bf r}-{\bf r}'}{2},\nonumber\\
&&\bar t\equiv t-t',\qquad T\equiv \frac{t+t'}{2}.
\end{eqnarray}
Here the relative coordinates, $(\bar\r,\bar t)$, describe the microscopic ``fast" dynamics and are treated exactly, while the center-of-mass coordinates, $(\R,T)$, describe macroscopic ``slow" dynamics and are treated semiclassically.
In order to separate out into the variables describing ``slow'' and ``fast'' processes,
we introduce the Wigner representation, which is defined by the Fourier transforms of the relative coordinates
\begin{eqnarray}
G(\bar\r,\bar t;{\bf R},T)=\int\frac{d{\bf k}}{(2\pi)^3}\int\frac{d\omega}{2\pi}\ e^{i({\bf k}\cdot\bar\r-\omega \bar t)}G({\bf k},\omega;{\bf R},T).\label{2PI:Wigner}
\end{eqnarray}
After performing the approximation fro the condensate~(\ref{2PI:markovian}) and the gradient expansion for the noncondensate Green's functions through the
Wigner transformation~(\ref{2PI:Wigner}), Eq~(\ref{generalized_GP:derivative_of_Gamma2_HF}) becomes
\begin{eqnarray}
&&\frac{\delta \Gamma_2[{\bm \Phi},{\bm G}]}{\delta \Phi^*({\bf r},t)}\nonumber\\
&&=-\int_c dt' F({\bf r},t;{\bf r}',t')\Phi({\bf r}',t')\nonumber\\
&&=i\frac{g^2}{2}\int d{\bf r}'\int\frac{d{\bf k}_1}{(2\pi)^3}\frac{d{\bf k}_2}{(2\pi)^3}\frac{d{\bf k}_3}{(2\pi)^3}\int\frac{d\omega_1}{2\pi}\frac{d\omega_2}{2\pi}\frac{d\omega_3}{2\pi}\nonumber\\
&&{}\times e^{i({\bf k}_1+{\bf k}_2-{\bf k}_3-{\bf k}_c)\cdot\bar\r}\frac{1}{\omega_c-\omega_1-\omega_2+\omega_3+i\eta}\nonumber\\
&&{}\times\biggl\{
\left[G^>({\bf k}_1,\omega_1;{\bf R},T)-G^<({\bf k}_1,\omega_1;{\bf R},T)\right]G^<({\bf k}_2,\omega_2;{\bf R},T)\nonumber\\
&&{}\qquad\times G^<({\bf k}_3,\omega_3;{\bf R},T)\nonumber\\
&&{}\quad+G^<({\bf k}_1,\omega_1;{\bf R},T)\left[G^>({\bf k}_2,\omega_2;{\bf R},T)-G^<({\bf k}_2,\omega_2;{\bf R},T)\right]\nonumber\\
&&{}\qquad\times G^<({\bf k}_3,\omega_3;{\bf R},T)\nonumber\\
&&{}\quad+
\left[G^>({\bf k}_1,\omega_1;{\bf R},T)-G^<({\bf k}_1,\omega_1;{\bf R},T)\right]\nonumber\\
&&{}\qquad\times\left[G^>({\bf k}_2,\omega_2;{\bf R},T)-G^<({\bf k}_2,\omega_2;{\bf R},T)\right]G^<({\bf k}_3,\omega_3;{\bf R},T)\nonumber\\
&&{}\quad-
G^<({\bf k}_1,\omega_1;{\bf R},T)G^<({\bf k}_2,\omega_2;{\bf R},T)\nonumber\\
&&{}\qquad\times\left[G^>({\bf k}_3,\omega_3;{\bf R},T)-G^<({\bf k}_3,\omega_3;{\bf R},T)\right]\biggl\}\Phi({\bf R},T).\label{generalized_GP:gradient_Gamma_2}
\end{eqnarray}
We now define the non-equilibrium spectral function:
\begin{eqnarray}
A({\bf k},\omega;{\bf R},T)&\equiv& i\left[G^>({\bf k},\omega;{\bf R},T)-G^<({\bf k},\omega;{\bf R},T)\right]\nonumber\\
&=&-2{\rm Im}G^{R}({\bf k},\omega;{\bf R},T)\label{2PI:nonequilibrium_spectral}.
\end{eqnarray}
When we introduce a new unknown function $f({\bf k},\omega;{\bf R},T)$ by~\cite{kadanoff1963}
\begin{eqnarray}
&&i G^<({\bf k},\omega;{\bf R},T)\equiv A({\bf k},\omega;{\bf R},T)f({\bf k},\omega;{\bf R},T),\label{generalized_GP:lesser_Green}\\
&&i G^>({\bf k},\omega;{\bf R},T)\equiv A({\bf k},\omega;{\bf R},T)\left[1+f({\bf k},\omega;{\bf R},T)\right],\label{generalized_GP:greater_Green}
\end{eqnarray}
the relation Eq.~(\ref{2PI:nonequilibrium_spectral}) is satisfied. 
The non-equilibrium spectral function can be obtained from the equation of motion for the retarded Green's function, $G^{R}$.
In the quasi-particle approximation, the non-equilibrium spectral function is given by~\cite{kadanoff1962}
\begin{eqnarray}
A({\bf k},\omega;{\bf R},T)\simeq 2\pi\delta(\omega-\tilde\epsilon({\bf k};{\bf R},T)/\hbar),\label{generalized_GP:quasiparticle_spectral}
\end{eqnarray}
where the Hartree-Fock spectrum for the noncondensate is defined by
\begin{eqnarray}
\tilde\epsilon({\bf k};{\bf R},T)=\frac{\hbar k^2}{2m}+2g[n_c({\bf R},T)+\tn({\bf R},T)]+V_{\rm ext}({\bf R}),\label{generalized_GP:Hartree-Fock_energy}
\end{eqnarray}
with the noncondensate density being defined by $\tn({\bf r},t)\equiv\langle\tpsi^*({\bf r},t)\tpsi({\bf r},t)\rangle$. 
In the quasi-particle approximation~(\ref{generalized_GP:quasiparticle_spectral}),
the unknown function~$f$ is found to be  equivalent to the Wigner distribution function, which is defined by 
\begin{eqnarray}
f_{W}({\bf k},{\bf R},T)\equiv\int \frac{d\omega}{2\pi}iG^<({\bf k},\omega;{\bf R},T),\label{2PI:Wigner_function}
\end{eqnarray}
where
\be
&&iG^{<}(\k,\omega;\R,T)\nonumber\\
&&{}\equiv \int dt e^{-i\omega t}\int d\r e^{i\k\cdot\r}\ \langle\tilde\psi^{*}\left(\R+\frac{\r}{2},T+\frac{t}{2}\right)\tilde\psi\left(\R-\frac{\r}{2},T-\frac{t}{2}\right)\rangle.
\ee
The Wigner function is a quantum counterpart of the classical phase-space distribution function.
Using Eqs.~(\ref{generalized_GP:lesser_Green})$\sim$(\ref{generalized_GP:Hartree-Fock_energy}) in Eq.~(\ref{generalized_GP:gradient_Gamma_2}), one obtains the generalized GP equation~\cite{zaremba1999} 
\begin{eqnarray}
i\hbar\frac{\partial}{\partial t}\Phi({\bf r},t)=\biggl[
-\frac{\hbar^2}{2m}\nabla^2+V_{\rm ext}({\bf r})+gn_c({\bf r},t)
+2g\tn({\bf r},t)-i \hbar R({\bf r},t)\biggl]\Phi({\bf r},t),\nonumber\\
\label{dissipative:dissipative_GP}
\end{eqnarray}
where
\begin{eqnarray}
R({\bf r},t)&\equiv&2g^2\left(\frac{2\pi}{\hbar}\right)^4\int\frac{d{\bf k}_1}{(2\pi)^3}\frac{d{\bf k}_2}{(2\pi)^3}\frac{d{\bf k}_3}{(2\pi)^3}\nonumber\\
&&{}\times\delta(\omega_c+\omega_1-\omega_2-\omega_3)
\delta({\bf k}_c+{\bf k}_1-{\bf k}_2-{\bf k}_3)
\nonumber\\
&&{}\times\biggl\{[1+f({\bf k}_1,{\bf r},t)][1+f({\bf k}_2,{\bf r},t)]f({\bf k}_3,{\bf r},t)\nonumber\\
&&{}-f({\bf k}_1,{\bf r},t)f({\bf k}_2,{\bf r},t)[1+f({\bf k}_3,{\bf r},t)]\biggl\},\label{genralized_GP:dissipative_term}
\end{eqnarray}
with ${\bf p}_c=\hbar{\bf k}_c\equiv\hbar\nabla\theta$, $\epsilon_c=\hbar\omega_c\equiv-\hbar\partial_t\theta$, and $\tilde\epsilon({\bf k}_i)=\hbar\omega_i$ being the condensate momentum, condensate energy and thermal cloud energy, respectively.  
We note that Eq.~(\ref{dissipative:dissipative_GP}) is not a closed equation because of the noncondensate mean field
$\tilde n({\bf r},t)\equiv\langle\tilde \psi^*({\bf r},t)\tilde \psi({\bf r},t)\rangle$ and the non-equilibrium distribution function
$f({\bf k},{\bf r},t)$ in $R({\bf r},t)$.
One should solve coupled equations that consist of the generalized GP equation and some kind of a kinetic equation, which is derived from the non-equilibrium Dyson equation~(\ref{2PI:Dyson_Beliaev}), for the non-equilibrium distribution function $f({\bf k},{\bf r},t)$.~\cite{zaremba1999}

One can discuss collective modes of the condensate in the presence of the noncondensate atoms using only 
the generalized GP equation~(\ref{dissipative:dissipative_GP}) as long as
noncondensate atoms are in static equilibrium,~\cite{williams2001_1,williams2001_2,duine2001} or can be treated within the linear response theory.
In the former case, the non-equilibrium distribution function in the dissipative term $R({\bf r},t)$ is replaced with the static equilibrium Bose distribution function. On the other hand, in the latter case, the dissipative term is neglected and only the mean field is considered by the linear response theory.
In the previous paper,~\cite{konabe2006_2} we studied the microscopic mechanism of the Landau instability in a one-dimensional optical lattice
using the generalized GP equation and the associated Bogoliubov-de Genne equations derived from the Bose-Hubbard Hamiltonian with the
static equilibrium approximation for the noncondensate atoms.

\subsection{Generalized GP Hydrodynamic Equations for the Condensate}\label{subsec:generalized_hyd}
In order to study long-wavelength excitations, it is more convenient to use the hydrodynamic formulation.
From the generalized GP action, $S_{\rm GGP}[{\bm\Phi},{\bm G}]$, which obtained in the previous subsection, one can derive an action in terms of the density and the phase of the condensate order parameter by using the following variable transformation:
\begin{eqnarray}
\Phi({\bf r},t)=\sqrt{n_c({\bf r},t)}e^{i\theta({\bf r},t)},
\end{eqnarray}
where $n_c({\bf r},t)$ and $\theta({\bf r},t)$ are the density and  the phase of the condensate, respectively. 
The generalized GP hydrodynamic action is given as follows:
\be
&&S_{\rm GGP}[n_c,\theta,G]\nonumber\\
&&{}=\int d{\bf r}\int_c dt\ \biggl[\frac{i\hbar}{2}\frac{\partial n_c({\bf r},t)}{\partial t}-\hbar n_c({\bf r},t)\frac{\partial \theta({\bf r},t)}{\partial t}\biggl]\nonumber\\
&&{}+\int d{\bf r}\int_c dt\ \biggl\{
\frac{\hbar^2}{2m}\sqrt{n_c({\bf r},t)}\nabla^2\sqrt{n_c({\bf r},t)}-\frac{\hbar^2}{2m}n_c({\bf r},t)
\left[\nabla\theta({\bf r},t)\right]^2\biggl\}\nonumber\\
&&{}-\int d{\bf r}\int_c dt\ n_c({\bf r},t)\left[V_{\rm ext}({\bf r},t)+\frac{g}{2}n_c(\r,t)+2g\tn(\r,t)\right]\nonumber\\
&&{}-\int d{\bf r} d{\bf r}'\int_c dtdt'\ \sqrt{n_c({\bf r},t)}\sqrt{n_c({\bf r}',t')}
 e^{-i[\theta({\bf r},t)-\theta({\bf r}',t')]} F({\bf r},t;{\bf r}',t'),
 \label{generalized_GP:hydrodynamic_action}
\ee
where $F$ is defined by Eq.~(\ref{generalized_GP:F}).
By minimizing the action~(\ref{generalized_GP:hydrodynamic_action}) with respect to the density and phase and by performing the gradient expansion as performed in the previous section,  one obtains the generalized GP hydrodynamic equations, which are equivalent to the generalized GP equation~(\ref{dissipative:dissipative_GP})~\cite{zaremba1999,williams2001_1}
\begin{eqnarray}
&&\frac{\partial n_c({\bf r},t)}{\partial t}+\nabla\cdot\left[n_c({\bf r},t){\bf v}_{c}({\bf r},t)\right]=-\Gamma_{12}({\bf r},t),\label{generalized_hyd:continuity_eq}\\
&&m\frac{\partial {\bf v}_c({\bf r},t)}{\partial t}+\nabla\biggl[
\mu_c+\frac{m}{2}{\bf v}_c^2({\bf r},t)\biggl]=0,\label{generalized_hyd:josephson_eq}
\end{eqnarray}
where $\Gamma_{12}({\bf r},t)\equiv 2n_c({\bf r},t)R({\bf r},t)$ and the condensate chemical potential $\mu_c({\bf r},t)$ is given by
\begin{eqnarray}
\mu_c({\bf r},t)=-\frac{\hbar^2}{2m}\frac{\nabla^2\sqrt{n_c({\bf r},t)}}{\sqrt{n_c({\bf r},t)}}+V_{\rm ext}({\bf r},t)+gn_c({\bf r},t)+2g\tn({\bf r},t).\label{generalized_hyd:chemical_potential}
\end{eqnarray}
The condensate velocity is defined by ${\bf v}\equiv \hbar\nabla\theta/m$.
The hydrodynamic equations~(\ref{generalized_hyd:continuity_eq}) and (\ref{generalized_hyd:josephson_eq}) in the Thomas-Fermi
approximation were used to discuss the damping of condensate collective oscillations in the harmonic trap potential at finite
temperatures in Ref.~\onlinecite{williams2001_1}.

\section{COARSE-GRAINED FINITE-TEMPERATURE THEORY IN OPTICAL LATTICES}\label{sec:generalized_hydrodynamic}
Several authors have derived the hydrodynamic equations for the Bose condensate at
zero temperature including the effect of a one-dimensional periodic lattice potential by focusing on the dynamics with length scale larger than a lattice spacing.~\cite{machholm2003,kramer2003,kramer2002}
Using the hydrodynamic equations, Kr\"{a}mer \textit{et.al}~\cite{kramer2003,kramer2002} calculated frequencies of condensate collective
oscillations. They have found that the frequency is renormalized through the effective mass due to the lattice 
potential.~\cite{kramer2003,kramer2002}
Their results are found to be in good agreement with the experimental data obtained in
Ref.~\onlinecite{cataliotti2001}.

In addition to the experiments close to $T=0$, such as in Refs.~\onlinecite{burger2001,fallani2004,fertig2005,cataliotti2001},
there have been very interesting experiments on the Bose condensate 
in optical lattices in the presence of the thermal cloud,
such as damping in collective oscillation and the breakdown of superfluidity.~\cite{ferlaino2002,sarlo2005}
Thus, it is tempting to derive finite-temperature hydrodynamic equations in the presence of the
periodic lattice potential. 
In this section, we give a derivation of finite-temperature hydrodynamic equations, which are generalization of the hydrodynamic equations derived by K\"{a}mer \textit{et. al}~\cite{kramer2002,kramer2003} for the condensate, including the effects of the one-dimensional optical lattice
at finite temperatures

\subsection{Coarse-grained action}\label{subsec:coarse-grained-action}
In this subsection, we derive the coarse-grained action for Bose gases in optical lattice. 
Because the difficulty comes from the $z$-direction trap potential, we first consider only a lattice potential in the $z$-direction with a confining trap potential in the $\r_{\bot}$ directions. 
After that, the trap potential in the $z$-direction can be included by the local density approximation.
For this purpose, it is convenient to start with the generalized GP action $\Gamma_{{\bm \Phi}}[{\bm \Phi},{\bm G}]=S_{\rm GGP}[{\bm \Phi},{\bm G}]$, defined in Eq.~(\ref{2PI:Gamma_phi}), rather than the action defined in Eq.~(\ref{generalized_GP:hydrodynamic_action}).
Splitting this action into the three parts, one obtains
\begin{eqnarray}
S_{\rm GGP}[{\bm \Phi},{\bm G}]=S_1+S_2+S_3,\label{coarse-grained-action:action_S}
\end{eqnarray}
where
\begin{eqnarray}
S_1&=&-\int_{-d/2}^{d/2} dz\int d{\bf r}_{\bot}\int_c dt
\Phi^*({\bf r},t)\biggl[
-\frac{\hbar^2}{2m}\frac{\partial^2}{\partial z^2}+V_{\rm opt}(z)+\frac{g}{2}|\Phi({\bf r},t)|^2\nonumber\\
&&{}+2g\tn({\bf r},t)\biggl]\Phi({\bf r},t),\label{coarse-grained-action:S_1}\\
S_2&=&-\int dz\int d{\bf r}_{\bot}\int_c dt\Phi^*({\bf r},t)\biggl[-i\hbar\frac{\partial}{\partial t}-\frac{\hbar^2}{2m}\nabla_{\bot}^2
+V_{\rm ho}({\bf r}_{\bot})\biggl]\Phi({\bf r},t),\label{coarse-grained-action:S_2}\\
S_3&=&-\int dzdz'\int d{\bf r}_{\bot}d{\bf r}'_{\bot}\int_c dtdt'
\Phi^*({\bf r},t)F({\bf r},t;{\bf r}',t')\Phi({\bf r}',t').\label{coarse-grained-action:S_3}
\end{eqnarray}
Here ${\bf r}_{\bot}\equiv (x,y)$ and $\nabla_{\bot}\equiv \partial^2/\partial x^2+\partial^2/\partial y^2$. $V_{\rm opt}(z)$ is the optical lattice potential, while $V_{\rm ho}({\bf r})$ is the harmonic potential in Eq.~(\ref{2PI:potential}).
The function $F({\bf r},t;{\bf r}',t')$ involved in the nonlocal part $S_3$ is defined by Eq.~(\ref{generalized_GP:F}). 
The advantage of working with the action defined by Eqs.(\ref{coarse-grained-action:S_1})-(\ref{coarse-grained-action:S_3}) is that the terms involve integrations over position and thus one can introduce coarse-grained approximation in a well-defined manner.
We note that, in Eq.~(\ref{coarse-grained-action:S_1}), the domain of the integration in the $z$-direction is $-d/2\le z< d/2$ since we consider only the lattice potential in the $z$-direction.

For the variational parameter $\Phi(\r,t)$, we use the following {\it ansatz}
\be
\Phi(\r,t)=\Phi_{k_c}(z)\phi_c(\r_{\bot},t),\label{coarse-grained-action:bloch_condensate}
\ee
where $\Phi_{k_c}(z)=e^{ik_cz}u_{k_c}(z)$ is the Bloch function with a quasi-momentum $k_c$ in a lowest band. The condensate Bloch amplitude $u_{k_c}(z)$ has the periodicity of the lattice potential. This {\it ansatz} is exact for $V_{\rm opt}(z) =0$ and $V_{\rm ho}(z)=0$, i.e., uniform in the $z$-direction.
Similarly, the field operator for the noncondensate is assumed to be expanded by the Bloch states:
\be
\tpsi(\r,t)=\sum_{n}\sum_{k}\tphi_{n,k}(z)\hat{\vphi}_{n,k}(\r_{\bot},t),\label{coarse-grained-action:bloch_noncondensate}
\ee
where $\hat{\vphi}_{n,k}(\r_{\bot})$ is a destruction operator of the noncondensate in the $\r_{\bot}$-direction. The Bloch function $\tphi_{n,k}(z)=e^{ikz}\tu_{n,k}(z)$ describing the thermal cloud atoms with a quasi-momentum in a $n$-band, which satisfies the following Schr\"{o}dinger equation
\be
\hat{\mathcal{H}}(z)\tphi_{n,k}(z)=\tilde\varepsilon^{(0)}_{n,k}\tphi_{n,k}(z),\label{coarse-grained-action:Schrodinger_eq}
\ee 
where
\be
\hat\mathcal{H}(z)\equiv-\frac{\hbar^2}{2m}\frac{\partial^2}{\partial z^2}+V_{\rm opt}(z).
\ee
The Bloch amplitudes of both condensate and thermal atoms are assumed orthonormal,
\begin{eqnarray}
\int_{-d/2}^{d/2}dz\ u_{k_c}^*(z)u_{k'_c}(z)=\delta_{k_c,k'_c},\label{coarse-grained-action:condensate normalization} \\
\int_{-d/2}^{d/2}dz\ \tu_{n,k}^*(z)\tu_{n',k'}(z)=\delta_{n,n'}\delta_{k,k'}.\label{coarse-grained-action:noncondensate normalization}
\end{eqnarray}

Substituting Eqs.~(\ref{coarse-grained-action:bloch_condensate}) and 
(\ref{coarse-grained-action:bloch_noncondensate}) into Eq.~(\ref{coarse-grained-action:S_1}), 
one obtains
\be
S_{1}&=&-\int_{-d/2}^{d/2}dz\int d\r_{\bot}\int_cdt\ \Phi_{k_c}^*(z)\vphi_c^*(\rb,t)\biggl[
-\frac{\hbar^2}{2m}\frac{\partial^2}{\partial z^2}+V_{\rm opt}(z)\nonumber\\
&&{}+\frac{g}{2}\left|\Phi_{k_c}(z)\right|^2\left|\vphi_c(\rb,t)\right|^2\nonumber\\
&&{}+2g\sum_{n,n'}\sum_{k,k'}\tphi_{n,k}^*(z)\tphi_{n',k'}(z)\langle\hat\vphi_{n,k}^{\dag}(\rb,t)\hat\vphi_{n',k'}(\rb,t)\rangle\biggl]\Phi_{k_c}(z)\vphi_c(\rb,t)\nonumber\\
&=&-\int d\rb\int_cdt\ \left|\vphi_c(\rb,t)\right|^2\nonumber\\
&&{}\times\int_{-d/2}^{d/2}dz\ \Phi_{k_c}^*(z)\biggl[
-\frac{\hbar^2}{2m}\frac{\partial^2}{\partial z^2}+V_{\rm opt}(z)+\frac{g}{2}\left|\Phi_{k_c}(z)\right|^2\left|\vphi_c(\rb,t)\right|^2\nonumber\\
&&{}+2g\sum_{n,n'}\sum_{k,k'}\tphi_{n,k}^*(z)\tphi_{n',k'}(z)\langle\hat\vphi_{n,k}^{\dag}(\rb,t)\hat\vphi_{n',k'}(\rb,t)\rangle\biggl]\Phi_{k_c}(z).\label{coarse-grained-action:S_1 Bloch function}
\ee
Using the Bloch amplitude $u_{k_c}(z)$ and $\tu_{k}(z)$, Eq.~(\ref{coarse-grained-action:S_1 Bloch function}) becomes
\be
S_1&=&-\int d\rb\int_cdt\ \left|\vphi_c(\rb,t)\right|^2\nonumber\\
&&{}\times\int_{-d/2}^{d/2}dz\ 
u_{k_c}^*(z)\biggl[
-\frac{\hbar^2}{2m}\left(\frac{\partial}{\partial z}+ik_c\right)^2+V_{\rm opt}(z)\nonumber\\
&&{}+\frac{g}{2}\left|u_{k_c}(z)\right|^2\left|\vphi_c(\rb,t)\right|^2\nonumber\\
&&{}+2g\sum_{n,n'}\sum_{k,k'}\tu_{n,k}^*(z)\tu_{n',k'}(z)\langle\hat\vphi_{n,k}^{\dag}(\rb,z,t)\hat\vphi_{n',k'}(\rb,z,t)\rangle\biggl]u_{k_c}(z),\label{coarse-grained-action:Bloch S_1}
\ee
where $\hat\vphi_k(\rb,z,t)\equiv e^{ikz}\hat\vphi_k(\rb,t)$. 
Similarly, Eq.~(\ref{coarse-grained-action:S_2}) becomes
\be
S_2=-\int d\rb\int_c dt\ \vphi_c^*(\rb,t)\left[-i\hbar\frac{\partial}{\partial t}-\frac{\hbar^2}{2m}\nabla_{\bot}^2+V_{\rm ho}(\rb)\right]\vphi_c(\rb,t),\label{coarse-grained-action:Bloch S_2}
\ee
where we have used the normalization conditions, Eqs.~(\ref{coarse-grained-action:condensate normalization}) and (\ref{coarse-grained-action:noncondensate normalization}).

Finally we consider $S_3$. In the system only with the harmonic potential, the Wigner transformation was performed by expanding the Green's function by a plane wave, as reviews in Sec.~\ref{sec:review}. On the other hand, in the system only with the one-dimensional periodic lattice potential, the Green's function should be expanded by the Bloch function, which has the periodicity of the lattice potential. The Green's function expanded by the Bloch function is given by
\be
iG(\r,t;\r',t')&=&\sum_{n,n'}\sum_{k,k'}\tphi_{n,k}(z)\tphi_{n',k'}^*(z')\langle{\rm T_c}[\hat\vphi_{n,k}(\rb,t)\hat\vphi^{\dag}_{n',k'}(\rb',t')]\rangle\nonumber\\
&=&\sum_{n,n'}\sum_{k,k'}\tu_{n,k}(z)\tu_{n',k'}^*(z')e^{ikz}e^{-ik'z'}\langle{\rm T_c}[\hat\vphi_{n,k}(\rb,t)\hat\vphi^{\dag}_{n',k'}(\rb',t')]\rangle.
\ee
The Bloch amplitudes $\tu_{n,k}(z)$ and $\tu_{n,k}^*(z)$ satisfy Eq.~(\ref{coarse-grained-action:Schrodinger_eq}). 
It will be convenient to define the following Green's functions:
\be
&&ig^{n,n'}_{k,k'}(\rb,t;\rb',t')\equiv\langle{\rm T_c}[\hat\vphi_{n,k}(\rb,t)\hat\vphi^{\dag}_{n',k'}(\rb',t')]\rangle,\label{coarse-grained-action:g}\\
&&iG^{n,n'}_{k,k'}(\rb,z,t;\rb',z',t')\equiv e^{ikz}e^{-ik'z'}\langle{\rm T_c}[\hat\vphi_{n,k}(\rb,t)\hat\vphi^{\dag}_{n',k'}(\rb',t')]\rangle.\label{coarse-grained-action:G_k}
\ee
In terms of $g_{k,k'}^{n,n'}$ or $G_{k,k'}^{n,n'}$, the Green's function $G(\r,t;\r',t')$ can be written as
\be
G(\r,t;\r',t')&=&\sum_{n,n'}\sum_{k,k'}\tphi_{n,k}(z)\tphi_{n',k'}^*(z')g^{n,n'}_{k,k'}(\rb,t;\rb',t')\\
&=&\sum_{n,n'}\sum_{k,k'}\tu_{n,k}(z)\tu^*_{n',k'}(z')G^{n,n'}_{k,k'}(\rb,z,t;\rb',z',t').
\ee
These Green's functions are useful when we apply the coarse-graining procedure.
With use of Eqs.~(\ref{coarse-grained-action:bloch_condensate}), (\ref{coarse-grained-action:g}), and (\ref{coarse-grained-action:G_k}) in (\ref{coarse-grained-action:S_3}), one obtains
\be
S_3&=&-\frac{g^2}{2}\sum_{n_1,n_1'}\sum_{n_2,n_2'}\sum_{n_3,n_3'}
\sum_{k_1,k_1'}\sum_{k_2,k_2'}\sum_{k_3,k_3'}\nonumber\\
&&{}\times\int_{-d/2}^{d/2}dz\tphi_{n_1,k_1}(z)\tphi_{n_2,k_2}(z)\tphi_{n_3',k_3'}^*(z)\Phi_{k_c}^*(z)\nonumber\\
&&{}\times\int_{-d/2}^{d/2}dz'\ \tphi_{n_1',k_1'}^*(z')\tphi_{n_2',k_2'}^*(z')\tphi_{n_3,k_3}(z')\Phi_{k_c}(z')\nonumber\\
&&{}\times\int d\rb d\rb'\int_cdtdt'\ \vphi_c^*(\rb,t)\vphi_c(\rb',t')\nonumber\\
&&{}\times g^{n_1,n_1'}_{k_1,k_1'}(\rb,t;\rb',t')g^{n_2,n_2'}_{k_2,k_2'}(\rb,t;\rb',t')g^{n_3,n_3'}_{k_3,k_3'}(\rb',t';\rb,t).\label{coarse-grained-action:Bloch S_3}
\ee

Now we include a confining trap potential in the $z$-direction.
As in the zero-temperature case, we are only interested in the dynamics on a length scale much longer than the lattice constant $d$, and thus the local density approximation can be used to treat the trap potential. In addition, we assume that one can use the Bloch function for describing the condensate on this length scale. 
We assume that the axial size of the condensate, $R_z$, in the confining trap potential is much larger than the lattice constant $d$.
The difficulty comes from the rapid variation of the length scale by the lattice potential.
In order to treat this length scales appropriately, we first regard the terms $e^{ik_cz}\vphi_c(\rb,t)$ and $e^{ikz}\hat\vphi_{n,k}(\rb,t)$ in Eqs.~(\ref{coarse-grained-action:bloch_condensate}) and (\ref{coarse-grained-action:bloch_noncondensate}) as almost constant in the length scale of the lattice constant $-d/2\le z<d/2$, although, which still depends on $z$. Then the variation due to the length scale of the confining trap potential occurs in the length scale much longer than the lattice constant. The condensate order parameter and the Green's function for the noncondensate are assumed to vary as the site index $l$:
\be
\Phi_c(\r,t)&\to& u_{k_c}(z)e^{ik_c^{l}(\rb,t)z}\vphi_c(\rb,l,t)\nonumber\\
&\equiv&u_{k_c}(z)\Phi_c(\rb,l,t),\label{coarse-grained-action:order parameter site}\\
G(\r,t;\r',t')&\to&\sum_{n,n'}\sum_{k,k'}\tu_{n,k}(z)\tu^*_{n',k'}(z')e^{ik^{l}(\rb,t)z}e^{-ik'^{l'}(\rb',t)z'}\nonumber\\
&&{}\times g_{k,k'}^{n,n'}(\rb,l,t;\rb',l',t')\nonumber\\
&\equiv&\sum_{n,n'}\sum_{k,k'}\tu_{n,k}(z)\tu^*_{n',k'}(z')G_{k,k'}^{n,n'}(\rb,l,t;\rb',l',t').\label{coarse-grained-action:Green's function site}
\ee
We can also define the site-represented amplitude and phase of the condensate order parameter:
\be
\Phi_c(\rb,l,t)=\sqrt{n_c(\rb,l,t)}e^{i\mathcal{S}(\rb,l,t)},\label{coarse-grained-action:amplitude phase site}
\ee
where
\be
n_c(\rb,l,t)&\equiv&|\Phi_c(\rb,l,t)|^2,\label{coarse-grained-action:density site}\\
\mathcal{S}(\rb,l,t)&\equiv& k_c(\rb,l,t)z+\vartheta(\rb,l,t).\label{coarse-grained-action:phase site}
\ee
In Eq.~(\ref{coarse-grained-action:phase site}), $\vartheta(\rb,l,t)$ is the phase of $\vphi_c(\rb,l,t)$.
By using Eq.~(\ref{coarse-grained-action:phase site}), a site-represented condensate velocity can be defined by
\be
v_z(\rb,l,t)&\equiv&\frac{\hbar}{m}\frac{\partial}{\partial z}\mathcal{S}(\rb,l,t)\nonumber\\
&=&\frac{\hbar}{m}k_c(\rb,l,t),\\
v_{x,y}(\rb,l,t)&\equiv&\frac{\hbar}{m}\nabla_{\bot}\mathcal{S}(\rb,l,t).
\ee
The above site-represented quantities can be understood by identifying these quantities as averaged one:
\begin{eqnarray}
&&n_{c}(\rb,l,t)\equiv\frac{1}{d}\int_{ld-d/2}^{ld+d/2}dz\ n_c({\bf r},t),\\
&&v_{c}(\rb,l,t)=\frac{1}{d}\int_{ld-d/2}^{ld+d/2}dz\ \frac{\hbar}{m}\frac{\partial}{\partial z}\theta({\bf r},t),
\end{eqnarray}
where $n_c({\bf r},t)=|\Phi({\bf r},t)|^2$ and $\theta(\r,t)$ are the condensate density and phase, respectively, which are the solution of the generalized GP equation~(\ref{dissipative:dissipative_GP}). 
One can also introduce the averaged Green's function
\begin{eqnarray}
G_{k,k'}^{n,n'}(\rb,l,t;\rb',l',t')\equiv\frac{1}{d^2}\int_{ld-d/2}^{ld+d/2}dz\int_{l'd-d/2}^{l'd+d/2}dz'\ G^{n,n'}_{k,k'}(\r,t;\r',t'),
\end{eqnarray}
where $G^{n,n'}_{k,k'}(\r,t;\r',t)$ is the solution of some kind of kinetic equation for the noncondensate atoms, for instance, the Kadanoff-Baym equations~\cite{kadanoff1962}. 

Substituting Eqs.~(\ref{coarse-grained-action:order parameter site}) and (\ref{coarse-grained-action:Green's function site}) into Eqs.~(\ref{coarse-grained-action:Bloch S_1}), (\ref{coarse-grained-action:Bloch S_2}), and (\ref{coarse-grained-action:Bloch S_3}), one obtains 
\be
S_1&=&-\sum_l\int d\rb\int_c dt\left|\Phi_c(\rb,l,t)\right|^2\int_{-d/2}^{d/2}dz\ u_{k_c}^*(z)\nonumber\\
&&{}\times\biggl[-\frac{\hbar^2}{2m}\left(\frac{\partial}{\partial z}+ik_c^{(l)}\right)^2+V_{\rm opt}(z)+\frac{g}{2}\left|u_{k_c}(z)\right|^2\left|\vphi_c(\rb,l,t)\right|^2\nonumber\\
&&{}+2g\sum_{n,n'}\sum_{k,k'}\tu_{n,k}^*(z)\tu_{n',k'}(z)iG_{k,k'}^{n,n'<}(\rb,l,t;\rb,l,t)\biggl]u_{k_c}(z),\label{coarse-grained-action:S_1 site}\\
S_2&=&-\sum_l\int d\rb\int_c dt\nonumber\\
&&{}\times\Phi_c^*(\rb,l,t)\left[-i\hbar\frac{\partial}{\partial t}-\frac{\hbar^2}{2m}\nabla_{\bot}^2+V_{\rm ho}(\rb,l)\right]\Phi_c(\rb,l,t),\label{coarse-grained-action:S_2 site}\\
S_3&=&-\frac{g^2}{2}\sum_{l,l'}\sum_{n_1,n_1'}\sum_{n_2,n_2'}\sum_{n_3,n_3'}
\sum_{k_1,k_1'}\sum_{k_2,k_2'}\sum_{k_3,k_3'}\nonumber\\
&&{}\times\int d\rb d\rb'\int_c dtdt'\ \Phi_c^*(\rb,l,t)\Phi_c(\rb',l',t')\nonumber\\
&&{}\times G_{k_1,k_1'}^{n_1,n_1'}(\rb,l,t;\rb',l',t')G_{k_2,k_2'}^{n_2,n_2'}(\rb,l,t;\rb',l',t')G_{k_3,k_3'}^{n_3,n_3'}(\rb',l',t';\rb,l,t)\nonumber\\
&&{}\times \int_{-d/2}^{d/2}dz\ \tu_{n_1,k_1}(z)\tu_{n_2,k_2}(z)\tu_{n_3',k_3'}^*(z)u_{k_c}^*(z)\nonumber\\
&&{}\times \int_{-d/2}^{d/2}dz'\ \tu_{n_1',k_1'}^*(z')\tu_{n_2',k_2'}^*(z')\tu_{n_3,k_3}(z')u_{k_c}(z'),\label{coarse-grained-action:S_3 site}
\ee
where $V_{\rm ho}(\rb,l)$ is a confining harmonic trap potential which labeled by $l$ in the $z$-direction.

The averaged quantities in Eqs.~(\ref{coarse-grained-action:S_1 site}), (\ref{coarse-grained-action:S_2 site}), and (\ref{coarse-grained-action:S_3 site}) are assumed to be smooth functions of ${\bf r}_{\bot}$ and to vary slowly with the
lattice site index $l$. Thus, the coarse-grained macroscopic densities and velocity can be obtained by 
replacing the discrete index $l$ with the continuous variable $z=ld$. This retains only the information on length scale much larger than the lattice spacing $d$ by focusing on the ``macroscopic" dynamics. Thus the ``microscopic" information shorter than the lattice constant is averaged out and only enters the modified local condensate energy given below. 
We define the coarse-grained quantities as follows:
\be
\Phi_c(\rb,l,t)&\to&\bar\Phi_c(\rb,z=ld,t),\label{coarse-grained-action:averaged order parameter}\\
n_c(\rb,l,t)&\to&\bar n_c(\rb,z=ld,t),\label{coarse-grained-action:averaged condensate density}\\
v_c(\rb,l,t)&\to&\bar v_c(\rb,z=ld,t), \label{coarse-grained-action:averaged velocity}\\
G_{k,k'}^{n,n'}(\rb,l,t;\rb',l',t')&\to&\bar G^{n,n'}_{k,k'}(\rb,z=ld,t;\rb',z'=l'd,t').\label{coarse-grained-action:averaged Green's function}
\ee 
The coarse-grained phase of the condensate is related with $\bar v_c(\r,t)$ by the following equation:
\be
\nabla\bar \mathcal{S}(\r,t)\equiv \frac{m}{\hbar}\bar v_c(\r,t).\label{coarse-graine-action:averaged phase}
\ee
Note that this equation gives the definition of the coarse-grained phase of the condensate.
 
With the coarse-grained quantities, we can obtain the coarse-grained action
\be
\bar S_{CG}=\bar S_1+\bar S_2+\bar S_3,\label{coarse-grained-action:coarse S}
\ee
where
\be
\bar S_1&=&-\int d\r\int_c dt\ \left|\bar\Phi_c(\r,t)\right|^2\varepsilon_{\rm opt}[k_c,\r,t],\label{coarse-grained-action:coarse S_1}\\
\bar S_2&=&-\int d\r \int_c dt\ \bar\Phi_c^*(\r,t)\left[-i\hbar\frac{\partial}{\partial t}-\frac{\hbar^2}{2m}\nabla_{\bot}^2+V_{\rm ho}(\r)\right]\bar\Phi_c(\r,t),\label{coarse-grained-action:coarse S_2}\\
\bar S_3&=&-\int d\r d\r'\int_c dtdt'\bar\Phi_c^*(\r,t)\bar F(\r,t;\r',t')\bar\Phi_c(\r',t').\label{coarse-grained-action:coarse S_3}
\ee
Non local term $\bar F(\r,t;\r',t')$ in $S_3$ is given by
\be
\bar F(\r,t;\r',t')&\equiv&\frac{g^2}{2}\sum_{n_1,n_1'}\sum_{n_2,n_2'}\sum_{n_3,n_3'}
\sum_{k_1,k_1'}\sum_{k_2,k_2'}\sum_{k_2,k_2'}\nonumber\\
&&{}\times \bar G^{n_1,n_1'}_{k_1,k_1'}(\r,t;\r',t')\bar G^{n_2,n_2'}_{k_2,k_2'}(\r,t;\r',t')\bar G^{n_3,n_3'}_{k_3,k_3'}(\r',t';\r,t)\nonumber\\
&&{}\times \int_{-d/2}^{d/2}dz\ \tu_{n_1,k_1}(z)\tu_{n_2,k_2}(z)\tu_{n_3',k_3'}^*(z)u_{k_c}^*(z)\nonumber\\
&&{}\times \int_{-d/2}^{d/2}dz'\ \tu_{n_1',k_1'}^*(z')\tu_{n_2',k_2'}^*(z')\tu_{n_3,k_3}(z')u_{k_c}(z').\label{coarse-grained-action:coarse F}
\ee
In Eq.~(\ref{coarse-grained-action:coarse S_1}), we have defined
\be
\varepsilon_{\rm opt}[k_c,\r,t]&\equiv&\int_{-d/2}^{d/2}dz\ u_{k_c}^*(z)\biggl[-\frac{\hbar^2}{2m}\left(\frac{\partial}{\partial z}+ik_c\right)^2+V_{\rm opt}(z)\biggl]u_{k_c}(z)\nonumber\\
&&{}+\frac{g}{2}\bar n_c(\r,t)\int_{-d/2}^{d/2}|u_{k_c}(z)|^4\nonumber\\
&&{}+2g\sum_{n,n'}\sum_{k,k'}i\bar G_{k,k'}^{n,n'}(\r,t;\r,t)\int_{-d/2}^{d/2}dz\ |u_{k_c}(z)|^2\tu^*_{n,k}(z)\tu_{n',k'}(z).\label{coarse-grained-action:energy in lattice}
\ee

\subsection{Coarse-grained generalized GP equation}
In this subsection, we derive the coarse-grained GP equation from Eq.~(\ref{coarse-grained-action:coarse S}). 
By taking the derivative with respect to $\bar\Phi_c^*$, one obtains
\be
i\hbar\frac{\partial}{\partial t}\bar\Phi_c(\r,t)
&=&\biggl\{-\frac{\hbar^2}{2m}\nabla_{\bot}^2+\varepsilon_{\rm opt}(k_c,\r,t)
+V_{\rm ho}(\r)\biggl\}\bar\Phi_c(\r,t)\nonumber\\
&&{}+\int d\r'\int_c dt'\bar F(\r,t;\r',t')\bar\Phi_c(\r',t').
\ee
This equation involves the effects of the lattice potential by $\varepsilon_{\rm opt}(k_c,\r,t)$ and of the coupling to the thermal cloud by the non-local function $\bar F(\r,t;\r',t')$. The crucial point here is that the difficulty which comes from the rapid variation due to the lattice potential disappeared and the effects of the lattice potential is effectively included.

In order to further simplify the correlation function $\bar F$, we assume that the coarse-grained macroscopic variables vary slowly in space and time compared to the spatial and temporal scale of a collision event. 
We can then approximate the condensate order parameter at $(\r',t')$ close to $(\r,t)$ by a Taylor expansion
\be
\bar\Phi(\r',t')&=&\sqrt{\bn_c(\r',t')}e^{i\bar\mathcal{S}(r',t')}\nonumber\\
&\simeq&\sqrt{\bn_c(\r,t)}e^{i[\bar\mathcal{S}(\r,t)+\partial_t\bar\mathcal{S}(\r,t)(t'-t)
+\nabla\bar\mathcal{S}(\r,t)\cdot(\r'-\r)]}\nonumber\\
&\equiv&\bar\Phi(\r,t)e^{-i[\omega_c(\r,t)(t'-t)-\k_c(\r,t)\cdot(\r'-\r)]}.
\ee
The condensate frequency and wavevector are defined by $\omega_c(\r,t)=-\partial_t\bar\mathcal{S}(\r,t)$ and $\k_c(\r,t)=\nabla\bar\mathcal{S}(\r,t)$, respectively. For the Green's function, we follow the same procedure as in Sec.~\ref{subsec:generalized_GP} by using the Wigner transform. We note, however, that the Green's function $G_{k_{z1},k_{z2}}(\r_1,t_1;\r_2,t_2)$ in this section involves the band index $n$ and quasi-momentum $k_{z}$. 
We first expand the Green's function by the plane-wave and rewrite by the center-of-mass, $\R\equiv (\r_1+\r_2)/2$, and relative coordinates, $\r\equiv \r_1-\r_2$
\be
G_{k_{z1},k_{z2}}^{n,n'<}(\r_1,t_1;\r_2,t_2)
&=&e^{ik_{z1}z_1}e^{-ik_{z2}z_2}
\int\frac{d\k_{\bot,1}}{(2\pi)^2}
\int\frac{d\k_{\bot,2}}{(2\pi)^2}\nonumber\\
&&{}\times e^{i\k_{\bot1}\cdot\r_{\bot1}}
e^{-i\k_{\bot2}\cdot\r_{\bot2}}
G^{n,n'<}(\k_1,\k_2,t_1,t_2)\nonumber\\
&\simeq&\delta_{n,n'}\ e^{iK_zz}\int\frac{d\K_{\bot}}{(2\pi)^2}
e^{i\K_{\bot}\cdot\rb}
\int\frac{d\k_{\bot}}{(2\pi)^2}\nonumber\\
&&{}\times e^{ik_zZ}e^{i\k_{\bot}\cdot\R_{\bot}}G_n^{<}(\k,\K,t,t'),
\ee
where we have introduced $\K\equiv(\k_1+\k_2)/2$ and $\k\equiv\k_1-\k_2$ and neglect the multi-band effects in the Green's function. We have used a notation: $G_n^{<}(\k,\K,t,t')\equiv G^{n,n<}(\k,\K,t,t')$.
The Wigner transformed Green's function is defined by
\be
G_n^<(\K,\omega;\R,T)&\equiv&\int\frac{d\omega}{2\pi}e^{-i\omega(t-t')}G_n^<(\K,\R,t,t')\nonumber\\
&\equiv&\int\frac{d\omega}{2\pi}e^{-i\omega(t-t')}\int\frac{d\k}{(2\pi)^3}e^{i\k\cdot\R}G_n^<(\k,\K,t,t'),\label{coarse-grained-GP: Wigner Green's function}
\ee
where the integral for the quasi-momentum $k_z$ comes from the one in Eq.~(\ref{coarse-grained-action:coarse F}).
We then use the quasi-particle approximation, changing the notation $\K\to\k$
\be
&&iG_n^{<}(\k,\omega;\R,T)=2\pi\delta(\omega-\tilde\epsilon_n(\k;\R,T)/\hbar)f_n(\k,\R,T),\label{coarse-grained-GP:quasi-particle G^<}\\
&&iG_n^{>}(\k,\omega;\R,T)=2\pi\delta(\omega-\tilde\epsilon_n(\k;\R,T)/\hbar)[1+f_n(\k,\R,T)],\label{coarse-grained-GP:quasi-particle G^>}
\ee
where the noncondensate energy $\tilde\epsilon$ is derived from the equation of motion for the retarded Green's function and is given by
\be
\tilde\epsilon_n(\k;\R,T)&\equiv&\frac{\hbar^2\k^2_{\bot}}{2m}+\tilde\varepsilon_{n,k_z}^{(0)}+2g\bar n_c(\R,T)
\int_{-d/2}^{d/2}dz\ |u_{k_c}(z)|^2|\tu_{n,k_z}(z)|^2\nonumber\\
&&{}+2g\sum_{m}\int\frac{d\q}{(2\pi)^3}f_m(\q;\R,T)\int_{-d/2}^{d/2}dz\ |\tu_{m,q_z}(z)|^2|\tu_{n,k_z}(z)|^2,\label{coarse-grained-GP:noncondensate energy}
\ee
where $\tilde\varepsilon_{n,k_z}^{(0)}$ is a solution of Eq.~(\ref{coarse-grained-action:Schrodinger_eq}).
With use of the Wigner transformed Green's function (\ref{coarse-grained-GP: Wigner Green's function}) and the quasi-particle approximation (\ref{coarse-grained-GP:quasi-particle G^<}), Eq.~(\ref{coarse-grained-action:energy in lattice}) becomes
\be
\varepsilon_{\rm opt}(k_c,\R,T)
&\equiv&\int_{-d/2}^{d/2}dz\ u_{k_c}^*(z)\biggl[-\frac{\hbar^2}{2m}\left(\frac{\partial}{\partial z}+ik_c\right)^2+V_{\rm opt}(z)\biggl]u_{k_c}(z)\nonumber\\
&&{}+\frac{g}{2}\bar n_c(\R,T)\int_{-d/2}^{d/2}dz\ |u_{k_c}(z)|^4\nonumber\\
&&{}+2g\sum_{n}\int\frac{d\k}{(2\pi)^3}f_{n}(\k;\R,T)\int_{-d/2}^{d/2}dz\ |\tu_{n,k_z}(z)|^2|u_{k_c}(z)|^2.\label{coarse-grained-GP:energy in lattice}
\ee

With these approximations, we can derive the generalized GP equation in an optical lattice potential, replacing $(\R,T)$ with $(\r,t)$
\be
i\hbar\frac{\partial}{\partial t}\bar\Phi_c(\r,t)
&=&\biggl\{-\frac{\hbar^2}{2m}\nabla_{\bot}^2+\varepsilon_{\rm opt}(k_c,\r,t)
+V_{\rm ho}(\r)+i\bar R(\r,t)\biggl\}\bar\Phi_c(\r,t).\label{coarse-grained-GP:coarse GP}
\ee
where the dissipative term $\bar R(\r,t)$ is given by
\be
\bar R(\r,t)&\equiv&2\left(\frac{2\pi}{\hbar}\right)^4\sum_{m\in{\bf N}}\sum_{n_1,n_2,n_3}\int\frac{d\k_1}{(2\pi)^3}\frac{d\k_2}{(2\pi)^3}\frac{d\k_3}{(2\pi)^3}\nonumber\\
&&{}\times\left| g\int_{-d/2}^{d/2}dz u^*_{k_{z,c}}(z)\bu_{n_1,k_{z1}}(z)\bu_{n_2,k_{z2}}(z)\bu^*_{n_3,k_{z3}}(z)\right|^2\nonumber\\
&&{}\times\delta(\omega_c+\omega_1-\omega_2-\omega_3)\nonumber\\
&&{}\times\delta(\k_{\bot c}+\k_{\bot1}-\k_{\bot2}-\k_{\bot3})\nonumber\\
&&{}\times\delta(k_{zc}+k_{z1}-k_{z2}-k_{z3}-2mq_{B})\nonumber\\
&&{}\times\biggl\{\left[1+f_{n_1}(\k_1,\r,t)\right]\left[1+f_{n_2}(\k_2,\r,t)\right]f_{n_3}(\k_3,\r,t)\nonumber\\
&&{}\quad\quad-f_{n_1}(\k_1,\r,t)f_{n_2}(\k_2,\r,t)\left[1+f_{n_3}(\k_3,\r,t)\right]\biggl\}.\label{hydrodynamic:dissipative}
\ee
Here ${\bf p}_c=\hbar{\bf k}_c$ is the condensate momentum and $q_B=\pi/d$ in the umklapp term $2mq_B$ (where $m$ is an integer) in the $\delta$-function is the Bragg wave-number . Compared Eq.~(\ref{hydrodynamic:dissipative}) with Eq.~(\ref{genralized_GP:dissipative_term}), we see two new features associated with the lattice potential. First, the bare coupling constant $g$ is renormalized by the Bloch functions. Secondly, the momentum conservation for the $z$-direction is modified to the looser condition: $\hbar k_{z,c}+\hbar k_{z,1}-\hbar k_{z,2}-\hbar k_{z,3}=2m\hbar q_B$. This reflects the breakdown of the translational symmetry due to the periodic optical lattice potential. 
In the formalism developed here, the collision dynamics information of a length scale shorter than the lattice spacing $d$ is effectively included through these two new features. 
The local condensate energy, $\epsilon_c\equiv \hbar\omega_c$, is defined by
\be
\epsilon_c(\r,t)\equiv -\frac{\hbar^2}{2m}\frac{\nabla_{\bot}^2\sqrt{\bn_c({\bf r},t)}}{\sqrt{\bn_c({\bf r},t)}}+\mu_{\rm opt}(k_c,\r,t)+V_{\rm ho}({\bf r})+\frac{m}{2}\bar{\bf v}_{c,\bot}^2,\label{hydrodynamic:condensate_energy}
\ee
where $\mu_{\rm opt}(k_c,\r,t)\equiv\partial(\bar n_c\varepsilon_{\rm opt})/\partial \bar n_c$. This expression for the condensate energy is given below after discussing the quantum hydrodynamic formulation for the condensate by the coarse-grained quantities.
The noncondensate energy, $\tilde\epsilon_i\equiv \hbar\omega_i$, in Eq.~(\ref{hydrodynamic:dissipative}) is defined by $\tilde\epsilon_i\equiv \tilde\epsilon_{n_i}(\k_i,\r,t)$.

\subsection{Coarse-grained generalized GP hydrodynamic equations}\label{subsec:hydrodynamic_equations}
In order to derive the coarse-grained equations in terms of hydrodynamic variables, we should start from the coarse-grained action because $\varepsilon_{\rm opt}$ in the coarse-grained generalized GP equation~(\ref{coarse-grained-GP:coarse GP}) depends on the condensate velocity through the wavevector $\k_c$, preventing usual variable transformation from $\bar \Phi_c$ and $\bar\Phi_c^*$ to $\bar n_c$ and $\bar \v_c$ in the coarse-grained generalized GP equation Eq.~(\ref{coarse-grained-GP:coarse GP}).

Combining the results given in Eqs.~(\ref{coarse-grained-action:coarse S_1}), (\ref{coarse-grained-action:coarse S_2}), and (\ref{coarse-grained-action:coarse S_3}) with Eq.~(\ref{coarse-grained-action:energy in lattice}), and using the hydrodynamic variables, Eqs.~(\ref{coarse-grained-action:averaged condensate density}), (\ref{coarse-graine-action:averaged phase}), we can write down the effective action for the condensate in optical lattices in terms of the coarse-grained macroscopic variables
\begin{eqnarray}
\bar S_{\rm eff}&=&\int d{\bf r}\int_c dt\ \biggl[\frac{i\hbar}{2}\frac{\partial \bar n_c({\bf r},t)}{\partial t}-\hbar \bar n_c({\bf r},t)\frac{\partial \bar\mathcal{S}({\bf r},t)}{\partial t}\biggl]\nonumber\\
&&{}+\int d{\bf r}\int_c dt\ \biggl\{
\frac{\hbar^2}{2m}\sqrt{\bar n_c({\bf r},t)}\nabla_{\bot}^2\sqrt{\bar n_c({\bf r},t)}-\frac{\hbar^2}{2m}\bar n_c({\bf r},t)
\left[\nabla_{\bot}\bar\mathcal{S}({\bf r},t)\right]^2\biggl\}\nonumber\\
&&{}-\int d{\bf r}\int_c dt\ \bar n_c({\bf r},t)\varepsilon_{\rm opt}(k_c,\r,t)\nonumber\\
&&{}-\int d{\bf r}\int_c dt\ \bar n_c({\bf r},t)V_{\rm ho}({\bf r},t)\nonumber\\
&&{}-\int d{\bf r} d{\bf r}'\int_c dtdt'\ \sqrt{\bar n_c({\bf r},t)}\sqrt{\bar n_c({\bf r}',t')}
 e^{-i[\bar\mathcal{S}({\bf r},t)-\bar\mathcal{S}({\bf r}',t')]}\bar F({\bf r},t;{\bf r}',t').\nonumber\\
 \label{hydrodynamic:smoothed_action}
\end{eqnarray}
The quantum hydrodynamic equations for the condensate are found by minimizing the effective
action~(\ref{hydrodynamic:smoothed_action}) with respect to the density and phase, 
leading to two coupled equations:
\begin{eqnarray}
&&m\frac{\partial \bar{\bf v}_c({\bf r},t)}{\partial t}+\nabla\biggl[
-\frac{\hbar^2}{2m}\frac{\nabla_{\bot}^2\sqrt{\bar n_c({\bf r},t)}}{\sqrt{\bar n_c({\bf r},t)}}+\mu_{\rm opt}(k_c,\bar n_c,\bar{n}_{nc})+V_{\rm ho}({\bf r})+\frac{m}{2}\bar {\bf v}_{c,\bot}^2\biggl]\nonumber\\
&&{}=-\nabla\biggl\{{\rm Im}\biggl[\frac{1}{\sqrt{\bar n_c({\bf r},t)}}e^{-i\bar\mathcal{S}({\bf r},t)}\int d{\bf r}'\int_c dt\bar F({\bf r},t;{\bf r}',t')\sqrt{\bar n_c({\bf r}',t')}e^{i\bar\mathcal{S}({\bf r}',t')}\biggl]\biggl\},\nonumber\\
\label{hydrodynamic:Josepson_eq}\\
&&\frac{\partial \bar n_c({\bf r},t)}{\partial t}+\nabla_{\bot}\cdot\left[\bar n_c({\bf r},t)\bar{\bf v}_{c,\bot}({\bf r},t)\right]+\frac{1}{\hbar}\frac{\partial}{\partial z}
\left[\frac{\partial \varepsilon_{\rm opt}(k_c,\bar n_c,\bar{n}_{nc})}{\partial k_c}\bar n_c({\bf r},t)\right]\nonumber\\
&&{}=-2{\rm Im}\biggl[\sqrt{\bar n_c({\bf r},t)}e^{-i\bar\mathcal{S}({\bf r},t)}\int d{\bf r}'\int_c dt' \bar F({\bf r},t;{\bf r}',t')
\sqrt{\bar n_c({\bf r}',t')}e^{i\bar\mathcal{S}({\bf r}',t')}\biggl],\nonumber\\
\label{hydrodynamic:continuity_eq}
\end{eqnarray}
where we defined the condensate chemical potential
\begin{eqnarray}
\mu_{\rm opt}(k_c,\bar n_c,\bar n_{nc})=\frac{\partial \left(\bn_c\varepsilon_{\rm opt}(k_c,\bn_c,\bn_{nc})\right)}{\partial \bar n_c}.
\end{eqnarray}
The condensate equation of motion given by Eqs.~(\ref{hydrodynamic:Josepson_eq}) and (\ref{hydrodynamic:continuity_eq}) describe the long-wavelength dynamics of the condensate in the presence of the periodic lattice potential, including the crucial coupling to the thermal cloud described by the function $\bar F(\r,t;\r',t')$. All information associated with the length scale shorter than the lattice spacing $d$ is buried in the Bloch functions.

Similarly to the coarse-grained generalized GP equation, one can simplify the above equations by local approximations and obtain the following equations:
\be
m\frac{\partial \bar{\bf v}_c({\bf r},t)}{\partial t}&=&-\nabla\Bigg[
-\frac{\hbar^2}{2m}\frac{\nabla_{\bot}^2\sqrt{\bn_c({\bf r},t)}}{\sqrt{\bn_c({\bf r},t)}}+\mu_{\rm opt}(k_c,\bn_c,\bn_{nc})\nonumber\\
&&{}\qquad+V_{\rm ho}({\bf r})+\frac{m}{2}\bar{\bf v}_{c,\bot}^2\Bigg],
\label{hydrodynamic:Markovian_Josepson_eq}\\
\frac{\partial \bn_c({\bf r},t)}{\partial t}&=&-\nabla_{\bot}\cdot\left[\bn_c({\bf r},t)\bar{\bf v}_{c,\bot}({\bf r},t)\right]
-\frac{\partial}{\partial z}
\left[\frac{\partial \varepsilon_{\rm opt}(k_c,\bn_c,\bn_{nc})}{\hbar\partial k_c}\bn_c({\bf r},t)\right]\nonumber\\
&&{}-\bar{\Gamma}(\r,t),\nonumber\\
\label{hydrodynamic:Markovian_continuity_eq}
\ee 
where the source term in Eq.~(\ref{hydrodynamic:Markovian_continuity_eq}) is given by
\be
\bar{\Gamma}(\r,t)&\equiv& 2\bn_c(\r,t)\bar R(\r,t),\label{hydrodynamic:Gamma_12}
\ee
where $\bar R(\r,t)$ is defined by Eq.~(\ref{hydrodynamic:dissipative}).

At zero temperature, the GP hydrodynamic equations for the lattice potential have been derived  by the tight binding approximation in Ref.~\onlinecite{kramer2003}, where the equation of state for $\mu_{\rm opt}$ is assumed to have the same structure of the uniform system. Ref.~\onlinecite{machholm2003,taylor2003} have obtained the GP hydrodynamic equations for a general $\mu_{\rm opt}$ and $\varepsilon_{\rm opt}$.
Compared with the GP hydrodynamic equations derived in the above works, Eqs.~(\ref{hydrodynamic:Markovian_Josepson_eq}) and (\ref{hydrodynamic:Markovian_continuity_eq}) are finite-temperature generalization, which includes the interaction between the condensate and noncondensate atoms.  At $T=0$, of course, $\tn$ in $\mu_{\rm opt}$ and $\varepsilon_{\rm opt}$ and $\bar\Gamma$ vanishes, and thus Eqs.~(\ref{hydrodynamic:Markovian_Josepson_eq}) and (\ref{hydrodynamic:Markovian_continuity_eq}) reduce to the GP hydrodynamic equations for the lattice potential.

\subsubsection{Low condensate velocity approximation}
In this subsection, we restrict ourselves to a condensate moving with a small superfluid velocity. In that case, it is very useful to introduce the effective masses for the long wavelength in the lowest Bloch energy and chemical potential band and rewrite Eqs.~(\ref{hydrodynamic:Markovian_Josepson_eq}) and (\ref{hydrodynamic:Markovian_continuity_eq}) by using these effective mass.
The effective mass and chemical potential effective mass are defined by the following equations~\cite{machholm2003,kramer2003,taylor2003}:
\be
\frac{1}{m_0^*}&\equiv&\frac{\partial^2 \varepsilon_{\rm opt}(k_c)}{\hbar^2\partial k_c^2}\biggl|_{k_c=0},\label{hydrodynamic:effective_mass}\\
\frac{1}{m^*_{\mu,0}}&\equiv&\frac{\partial^2 \mu_{\rm opt}(k_c)}{\hbar^2\partial k_c^2}\biggl|_{k_c=0}.\label{hydrodynamic:chemical_effective_mass}
\ee
In the usual Bloch theory of metals, we have only one effective mass, namely, $m_0^*=m_{\mu,0}^*$. For the Bose condensate, however, due to the interaction term the two different effective masses enter the theory. This is originally comes from the two energies, the condensate energy band $\varepsilon_{\rm opt}$ and chemical potential band $\mu_{\rm opt}$.
In terms of these effective masses, (\ref{hydrodynamic:effective_mass}) and (\ref{hydrodynamic:chemical_effective_mass}), the coarse-grained generalized GP hydrodynamic equations, given by (\ref{hydrodynamic:Markovian_Josepson_eq}) and (\ref{hydrodynamic:Markovian_continuity_eq}), become
\be
&&m\frac{\partial \bar{\bf v}_c({\bf r},t)}{\partial t}+\nabla\biggl[
-\frac{\hbar^2}{2m}\frac{\nabla_{\bot}^2\sqrt{\bn_c({\bf r},t)}}{\sqrt{\bn_c({\bf r},t)}}+\mu_{\rm opt}(\bn_c,\bn_{nc})+V_{\rm ho}({\bf r})\nonumber\\
&&{}+\frac{1}{2}\left(\frac{m}{m^*_{\mu,0}}\right)m\bar v_z^2+\frac{m}{2}\bar{\bf v}_{c,\bot}^2\biggl]=0,\label{hydrodynamic:Markovian_small_currents_Josepson_eq}\\
&&\frac{\partial \bn_c({\bf r},t)}{\partial t}+\nabla_{\bot}\cdot\left[\bn_c({\bf r},t)\bar{\bf v}_{c,\bot}({\bf r},t)\right]
+\frac{\partial}{\partial z}\left[\left(\frac{m}{m^*_0}\right)v_{c,z}(\r,t)\bn_c(\r,t)\right]\nonumber\\
&&{}=-\bar{\Gamma}(\r,t),\label{hydrodynamic:Markovian_small_currents_continuity_eq}
\ee
where $\mu_{\rm opt}(\bn_c,\bn_{nc})\equiv \mu_{\rm opt}(k_c=0,\bn_c,\bn_{nc})$.
In the low condensate velocity approximation, the local condensate energy~(\ref{hydrodynamic:condensate_energy}) is given by
\be
\epsilon_c(\r,t)&\equiv&-\frac{\hbar^2}{2m}\frac{\nabla_{\bot}^2\sqrt{\bn_c({\bf r},t)}}{\sqrt{\bn_c({\bf r},t)}}+\mu_{\rm opt}(\bn_c,\bn_{nc})+V_{\rm ho}({\bf r})\nonumber\\
&&{}+\frac{1}{2}\left(\frac{m}{m^*_{\mu,0}}\right)m\bar v_z^2+\frac{m}{2}\bar{\bf v}_{c,\bot}^2.\label{hydrodynamic:small_current_condensate_energy}
\ee
The local condensate chemical potential is given by
\be
\mu_c(\r,t)\equiv\mu_{\rm opt}(\bn_c,\bn_{nc})+V_{\rm ho}({\bf r}).\label{hydrodynamic:small_current_chemical_potential}
\ee
When we set $\tilde \Gamma\!\to\! 0$ and $\mu_{\rm opt}(\bn_c,\bn_{nc})\!\to\!\mu_{\rm opt}(\bn_c,\bn_{nc}=0)$, appropriate when the thermal cloud is absent, Eqs.~(\ref{hydrodynamic:Markovian_small_currents_Josepson_eq}) and (\ref{hydrodynamic:Markovian_small_currents_continuity_eq}) reduce to those obtained  in Ref.~\onlinecite{kramer2003}.

\section{INSTABILITY OF THE SUPERFLUID}\label{sec:instability}
The Landau instability of the Bose condensate has been studied by using the original GP equation.~\cite{wu2001,wu2003,machholm2003,kramer2003,menotti2003,taylor2003}
Within the GP equation, one can derive the stability phase diagram for the condensate from the negative excitation energy condition. 
However, this approach does not give any insight into the microscopic origin of the Landau instability. Moreover, one has to construct a microscopic theory for the landau instability because the original Landau argument cannot be applied to the lattice system where the momentum conservation in not satisfied, which Landau originally relied on.

In the present section, we use the finite-temperature theory developed in the previous sections to exhibit a specific microscopic origin of 
the Landau instability of superfluidity in a Bose condensate.

\subsection{Relation between the damping of collective modes and the instability of the condensate}
We shall show that $\bar \Gamma$ defined by Eq.~(\ref{hydrodynamic:Gamma_12}) can change sign and as a result leads to a Landau instability.~\cite{konabe2006_2,iigaya2006,konabe2007_1} 
This is illustrated generally as follows.
The amplitude of the collective mode $\delta\Phi_{p_c,k}$ decays as 
$\delta\Phi_{p_c,k} \propto e^{-\bar\Gamma_{p_c,k}t}$, 
where $\bar\Gamma_{p_c,k}$ is a damping rate.
This relation indicates that the superfluid state is stable as long as $\bar\Gamma_{p_c,k}$ is positive for any collective mode momentum $k$.
In fact, for a condensate at rest ($p_c=0$) condensate, one can show that the damping rate is always
positive,~\cite{konabe2006_2,iigaya2006} and thus the collective mode decays exponentially in time.
However, as shown in the following subsections, $\bar\Gamma_{p_c,k}$ can become negative in the case of a moving condensate (finite $p_c$). A negative value of $\bar\Gamma_{p_c,k}$ indicates an instability. 
The exponential growth in time of the amplitude of collective modes due to collisional coupling to the thermal cloud destabilize the condensate state, leading to the breakdown of superfluidity. 
Recent experiment~\cite{sarlo2005} appear to support this scenario.

In the collisionless regime of interest to which we restrict ourselves, there are two important damping processes, namely, collisional damping and Landau damping. The former occurs due to the collisional exchange of atoms between the condensate and noncondensate,~\cite{williams2001_1,williams2001_2} while the latter occurs due to the dynamical coupling between
the condensate oscillation and the thermal excitations.~\cite{pitaevskii1997,fedichev1998,giorgini1998,fedichev1998_2,giorgini2000,tsuchiya2005}
In the present paper, we will only consider collisions between the condensate and noncondensate atoms, and also ignore the harmonic trap potential ($V_{\rm ho}(\r)=0$).

\subsection{Instability due to the Collisional Damping Process}\label{subsec:collisional}
The collisional damping process considered in the present paper was investigated by Williams and Griffin~\cite{williams2001_1,williams2001_2} and Duine and Stoof~\cite{duine2001} for collective modes of the condensate in the harmonic trap potential. The collisional damping arises due to the lack of diffusive equilibrium between the condensate and noncondensate, namely, equilibration process due to the collisional exchange between the condensate and the noncondensate. The authors of Refs.~\onlinecite{williams2001_1,williams2001_2,duine2001} calculated the collisional damping rate by solving the dynamical equation of motion for the condensate, while the noncondensate is treated as being static. 
To simplify the notation, we omit the bars in this section, however, we note that $n_c$, $\v_c$, and $\Gamma$ always refer to coarse-grained values.


We approximate the non-equilibrium distribution function in Eq.~(\ref{hydrodynamic:dissipative}) by the static Bose distribution function for thermal equilibrium~\cite{williams2001_1,williams2001_2,williams2002}
\begin{eqnarray}
f_n({\bf k})=\frac{1}{e^{\beta[\tilde{\epsilon}_n(\k)-\tilde{\mu}_0]}-1},
\end{eqnarray}
where $\tilde\mu_0$ is the chemical potential of the noncondensate atoms and 
the energy of the noncondensate atoms is given by the Hartree-Fock approximation appropriate to one-dimensioal lattice potential along the $z$-axis; $\tilde\epsilon_n(\k)\equiv \hbar\tilde{\omega}_n(\k)$, which is defined by Eq.~(\ref{coarse-grained-GP:noncondensate energy}).
Our use of the static thermal cloud approximation implicitly assumes that the thermal excitations reach equilibrium with a relaxation time much shorter than the period of the condensate collective modes. 
This assumption may be justified by the experiment~\cite{ferlaino2002}, where the thermal cloud in the lattice potential reaches its equilibrium state very rapidly.

Using the identity for the Bose distribution 
\begin{eqnarray}
f_1(1+f_2)(1+f_3)=(1+f_1)f_2f_3e^{\beta(\tilde{\epsilon}_{k_1}-\tilde{\epsilon}_{k_2}-\tilde{\epsilon}_{k_3}-\tilde{\mu}_0)},
\end{eqnarray}
the source term $\Gamma({\bf r},t)$ in Eq.~(\ref{hydrodynamic:Gamma_12}) reduce to
\begin{eqnarray}
\Gamma({\bf r},t)
&=&4n_c(\r,t)\left(\frac{2\pi}{\hbar}\right)^4\sum_{m\in {\bf N}}\sum_{n_1,n_2,n_3}\int\frac{d{\bf k}_1}{(2\pi)^3}\frac{d{\bf k}_2}{(2\pi)^3}\frac{d{\bf k}_3}{(2\pi)^3}\nonumber\\
&&{}\times\left| g\int_{-d/2}^{d/2}dz\ u_{k_{zc}}(z)\tu_{n_1,k_{z1}}(z)\bu_{n_2,k_{z2}}(z)\tu^*_{n_3,k_{z3}}(z)\right|^2\nonumber\\
&&{}\times\delta(\omega_c+\omega_1-\omega_2-\omega_3)\nonumber\\
&&{}\times\delta({\bf k}_{\bot c}+{\bf k}_{\bot 1}-{\bf k}_{\bot 2}-{\bf k}_{\bot 3})\nonumber\\
&&{}\times\delta(k_{zc}+k_{z1}-k_{z2}-k_{z3}-2mq_B)
\nonumber\\
&&{}\times\left[1-e^{\beta(\epsilon_c-\tilde{\mu})}\right]
[1+f_{n_1}({\bf k}_1)]f_{n_2}({\bf k}_2)f_{n_3}({\bf k}_3).\nonumber\\
\end{eqnarray}
We will derive linearized equations of Eqs.~(\ref{hydrodynamic:Markovian_Josepson_eq}) and (\ref{hydrodynamic:Markovian_continuity_eq}) with $n_c(\r,t)=n_c^0+\delta n(z)$ and $\v_c(\r,t)=[v_{zc}^0+\delta v_c(z)]\hat{\bf z}$, where $n_{c}^0$ and $v_{zc}^0$ are static values of the condensate density and velocity, respectively.
We need to evaluate $\Gamma$ to first order in the deviations away from equilibrium. 
The condensate energy~(\ref{hydrodynamic:small_current_condensate_energy}) is expanded as
\begin{eqnarray}
\epsilon_c&\simeq& \mu_{\rm opt}^0+m\frac{\partial\mu_{\rm opt}^0}{\hbar\partial k_c}\delta v_{zc}+\frac{\partial\mu_{\rm opt}^0}{\partial n_c}\delta n_c\nonumber\\
&=&\tilde\mu_0+m\frac{\partial\mu_{\rm opt}^0}{\hbar\partial k_c}\delta v_{zc}+\frac{\partial\mu_{\rm opt}^0}{\partial n_c}\delta n_c.
\label{collisional:expansion_epsilon}
\end{eqnarray}
where $\mu_{\rm opt}^0\equiv \mu_{\rm opt}(k_{c0},n_{c0},n_{nc,0})$, and $n_{c0}$ and $v_{c0}\equiv\hbar k_{c0}$ are the condensate density and velocity in equilibrium, respectively.
Here we have used the relation $\mu_{\rm opt}^0=\tilde\mu_0$.
Note that the fluctuation of the noncondensate is neglected because we use the static thermal cloud approximation. 

Using Eq.~(\ref{collisional:expansion_epsilon}), one can also expand $e^{\beta(\epsilon_c-\tilde{\mu}_0)}$ in Eq.~(\ref{hydrodynamic:Gamma_12}) to give
\begin{eqnarray}
&&\exp[\beta(\epsilon_c-\tilde{\mu}_0)]\nonumber\\
&&{}\simeq
\exp\left\{\beta\left[\frac{\partial \mu_{\rm opt}^0}{\partial n_c}\delta n_c+m\frac{\partial\mu_{\rm opt}^0}{\hbar\partial k_c}\delta v_{zc}\right]\right\}\nonumber\\
&&{}\simeq1+\beta\left[\frac{\partial \mu_{\rm opt}^0}{\partial n_c}\delta n_c+m\frac{\partial\mu_{\rm opt}^0}{\hbar\partial k_c}\delta v_{zc}\right].\nonumber\\
\end{eqnarray}
The resulting linearized form of the dissipation term $\Gamma({\bf r},t)$ reduces to
\begin{eqnarray}
\delta\Gamma(z,t)\equiv\frac{\beta n_c^0}{\tau}\biggl[\frac{\partial \mu_{\rm opt}^0}{\partial n_c}\delta n_c(z,t)+m\frac{\partial\mu_{\rm opt}^0}{\hbar\partial k_c}\delta v_{zc}(z,t)\biggl],\label{collisional:delta_Gamma}
\end{eqnarray}
where the relaxation time $\tau$ arising from the collisions between the condensate and noncondensate atoms is defined by
\begin{eqnarray}
\frac{1}{\tau}&\equiv&4\left(\frac{2\pi}{\hbar}\right)^4\sum_{m\in{\bf N}}
\sum_{n_1,n_2,n_3}
\int\frac{d\k_1}{(2\pi)^3}\frac{d\k_2}{(2\pi)^3}\frac{d\k_3}{(2\pi)^3}\nonumber\\
&&{}\times\left| g\int_{-d/2}^{d/2}dz\ u^*_{k_{zc}}(z)\tu_{n_1,k_{z1}}(z)\tu_{n_2,k_{z2}}(z)\tu^*_{n_3,k_{z3}}(\bz)\right|^2\nonumber\\
&&{}\times\delta(\omega_{c}^{0}+\omega_1^{0}-\omega_2^{0}-\omega_3^{0})\nonumber\\
&&{}\times\delta(\k_{\bot c}+\k_{\bot 1}-\k_{\bot 2}-\k_{\bot 3})\nonumber\\
&&{}\times\delta(k_{zc}+k_{z1}-k_{z2}-k_{z3}-2mq_{B})\nonumber\\
&&{}\times\left[1+f_{n_1}({\bf k}_1)\right]f_{n_2}({\bf k}_2)f_{n_3}({\bf k}_3),\label{collisional:relaxation_time}
\end{eqnarray}
where the superscript $``0"$ of $\omega_c^0$ and $\omega_i^0$ (where $i=1,2,3$) indicates the quantities take its static value.
We use Eq.~(\ref{collisional:delta_Gamma}) in the linearized version of the generalized GP hydrodynamic equations, leading to
\begin{eqnarray}
\frac{\partial \delta n_c(z,t)}{\partial t}&=&
-\frac{\partial}{\partial z}\biggl[\frac{m}{m^*}n_c^0\delta v_c(z,t)+v_c^{\mu}\delta n_c(z,t)\biggl]\nonumber\\
&&{}-\frac{\beta n_c^0}{\tau}\biggl[\frac{\partial \mu_{\rm opt}^0}{\partial n_c}\delta n_c(z,t)+mv_c^{\mu}\delta v_c(z,t)\biggl],
\label{collisional_linearized_continuity_eq}\\
m\frac{\partial}{\partial t}\delta{v}_c(z,t)&=&-\frac{\partial}{\partial z}\biggl[\frac{\partial\mu_{\rm opt}^0}{\partial n_c}\delta n_c(z,t)+mv_c^{\mu}\delta v_c(z,t)\biggl],
\label{collisional_linearized_josephson_eq}
\end{eqnarray}
where we have defined the effective mass and the chemical potential group velocity at the arbitrary value of the condensate velocity $k_c^0$ as follows~\cite{kramer2003,menotti2003}:
\be
\frac{1}{m^*}&\equiv&\frac{\partial^2\varepsilon_{\rm opt}}{\hbar^2\partial k_c^2}\biggl|_{k_c^0,n_c^0,n_{nc}^0},\label{collisional:effective mass}\\
v_c^{\mu}&\equiv&\frac{\partial \mu_{\rm opt}}{\hbar\partial k_c}\biggl|_{k_c^0,n_c^0,n_{nc}^0}\label{collisional:chemical potential group velocity}.
\ee
The effective mass $m^*$ is a finite-$k_c$ generalization of $m_0^*$ for the long wavelength defined by Eq.~(\ref{hydrodynamic:effective_mass}). 
We note that the chemical potential group velocity $v_c^{\mu}$ is deferent from the usual group velocity because $v_c^{\mu}$ is derived from the chemical potential band, while the usual group velocity is derived from the energy band $\varepsilon_{\rm opt}$. 

Before solving the coupled equations (\ref{collisional_linearized_continuity_eq}) and (\ref{collisional_linearized_josephson_eq}) for the condensate fluctuations, it is useful to derive the Stringari-type equation for the condensate fluctuation~\cite{stringari1996}, which gives the frequency of the condensate collective modes, in order to show the significance of the collisions between the condensate and noncondensate atoms buried in $\tau$.
One can show that the linearized equations (\ref{collisional_linearized_continuity_eq}) and (\ref{collisional_linearized_josephson_eq}) reduce to the finite-temperature Stringari equation in the presence of the lattice potential when we set $v_c^{\mu}=0$:
\be
\frac{\partial^2\delta n_c(z,t)}{\partial t^2}=\frac{n_c^0}{m^*}\frac{\partial^2}{\partial z^2}\left[\frac{\partial \mu_{\rm opt}^0}{\partial n_c}\delta n_c(z,t)\right]-\frac{1}{\tau'}\frac{\partial \delta n_c(z,t)}{\partial t},\label{collision:finite T Kramer}
\ee
where
\be
\frac{1}{\tau'}\equiv\frac{\beta n_c^0}{\tau}\frac{\mu_{\rm opt}^0}{\partial n_c}.
\ee
The finite-temperature Stringari equations was first derived by Williams and Griffin~\cite{williams2001_1} for a harmonic potential. 
The collision time $\tau'$ describes collisions between the condensate and noncondensate atoms when the condensate is perturbed away from equilibrium. Eq.~(\ref{collision:finite T Kramer}) clearly shows that the new term associated with the collision time $\tau'$ in Eq.~(\ref{collision:finite T Kramer}) causes damping of the condensate fluctuations. This damping is due to the lack of collisional detailed-balance between the condensate and the static thermal cloud pinned by the lattice potential. The collisional damping is important damping process in addition to the Landau damping in the collisionless regime.

To solve the coupled equations~(\ref{collisional_linearized_continuity_eq}) and (\ref{collisional_linearized_josephson_eq}), we assume a plane-wave solution $\sim{\rm exp}[i(q_zz-\omega t)]$ for both $\delta n_c$ and $\delta v_c$, and then Eqs.~(\ref{collisional_linearized_continuity_eq}) and (\ref{collisional_linearized_josephson_eq}) give
\begin{eqnarray}
&&\omega\delta n_c-\frac{m}{m^*}n_c^0q_z\delta v_c
-v_c^{\mu}q_z\delta n_c=i\frac{1}{\tau}\biggl(mv_c^{\mu}\delta v_c+\frac{\partial\mu_{\rm opt}^0}{\partial n_c}\delta n_c\biggl)
,\label{collisional:planewave_continuity_eq}\\
&&\omega\delta v_c=
\frac{1}{m}\frac{\partial \mu_{\rm opt}^0}{\partial n_c}q_z\delta n_c+v_c^{\mu}q_z\delta v_c
.\label{collisional:planewave_josephson_eq}
\end{eqnarray}
From Eq.~(\ref{collisional:planewave_josephson_eq}), one finds
\begin{eqnarray}
\delta v_c=\frac{1}{m}\frac{\partial \mu_{\rm opt}^0}{\partial n_c}q_z\frac{1}{\omega-v_c^{\mu}q_z}\delta n_c.
\end{eqnarray}
Substituting this into Eq.~(\ref{collisional:planewave_continuity_eq}) and eliminating $\delta v_c$, one obtains
\begin{eqnarray}
\left(\omega-v_c^{\mu}q_z\right)^2-\frac{n_c^0}{m^*}\frac{\partial \mu_{\rm opt}^0}{\partial n_c}q_z^2=i\frac{1}{\tau'}\omega.
\label{collisional:dispersion_eq}
\end{eqnarray}
In the absence of the collisions, i.e., taking the limit $1/\tau'\to 0$, we find the collective mode frequency $\Omega$ given by~\cite{machholm2003,kramer2003,menotti2003,taylor2003,kramer2002}
\begin{eqnarray}
\omega=v_{c}^{\mu}q_z\pm c^*q_z\equiv\Omega.\label{collisional:dispersion_relation}
\end{eqnarray}
Here
\begin{eqnarray}
c^*\equiv \sqrt{\frac{n_c^0}{m^*}\frac{\partial \mu_{\rm opt}^0}{\partial n_c}},\label{collisional:sound_velocity}
\end{eqnarray}
is a Bogoliubov-type sound velocity modified due to the presence of the lattice potential, while $v_{c}^{\mu}$ is defined by Eq.~(\ref{collisional:chemical potential group velocity}).
In Eq.~(\ref{collisional:dispersion_relation}), the opposite sign ``$\pm$" correspond to a sound wave propagating in the same and in the opposite direction, respectively. 
Kr\"{a}mer {\it et al.}~\cite{kramer2002} gave the sound velocity for an optical lattice by using tight-binding model with a specific approximation for the chemical potential. The general expression of the sound velocity was given by the GP hydrodynamic analysis by Machholm {\it et al.}~\cite{machholm2003} and by Kr\"{a}mer {\it et al.}~\cite{kramer2003}. Taylor and Zaremba used the Bogoliubov equation by a systematic expansion in powers of the phonon wave vector~\cite{taylor2003}. 
Our result of the sound velocity is natural extension of above works at zero temperature to finite temperature.
For a translationally invariant system and at zero temperature, $\mu_{\rm opt}\to gn_c^0$ and $m^*\to m$. Therefore, the sound velocity is given by the usual result $c=\sqrt{\frac{gn_c^0}{m}}$.

Without the collision term, the effect of the thermal cloud enter into the collective mode frequency by the mean-field interaction in the chemical potential. In this case, the collective mode does not damp. 
Now we study the effects of the collisions between the condensate and noncondensate atoms represented by the collision term $\tau'$, which is second order effect of the coupling constant. For this purpose, we include the effects of the collisions represented by the relaxation rate $1/\tau'$, giving the dispersion relation as $\omega=\Omega-i\Gamma_{\rm c}$. To first order in $1/\tau'$, one obtains the collisional damping rate to be
\begin{eqnarray}
\Gamma_{\rm c}=\frac{1}{2\tau'}\left(1\pm \frac{v_{c}^{\mu}}{c^*}\right)
.\label{collisional:damping_term}
\end{eqnarray}
This is the key relation to consider the instability of the condensate.
Recall the argument in the first part of this section. As far as the damping rate $\Gamma_c$ is positive, the collective mode of the condensate is stabilized by this damping process. This corresponds to the case of lower sign in Eq.~(\ref{collisional:damping_term}).
This expression, however, shows that $\Gamma_{\rm c}$ can be negative, indicating a growth instability when the direction of the condensate and sound velocity is opposite, and when
\begin{eqnarray}
v_{c}^{\mu}>c^*.
\end{eqnarray} 
This condition turns out to be same as the usual Landau criterion for the 
superfluidity in a uniform system, except that the condensate 
and the sound velocities are now modified due to the presence of the optical lattice
potential.
This type of instability has been discussed by imposing on the condition that the excitation energy becomes negative, i.e., $\hbar\Omega <0$. The region of this instability, so called Landau instability,   for an optical lattice potential was first obtained by Wu and Niu~\cite{wu2001,wu2003}. The same argument was performed by several authors~\cite{machholm2003,menotti2003,taylor2003,danshita2007}.
In contrast to the previous works, the crucial point in the present work, however, is that we 
derive the Landau criterion by specifying the microscopic destabilization process, which is performed by calculating the damping rate of the condensate collective mode. In this sense, we give a explanation of the microscopic mechanism of the Landau instability. 
This kind of discussion for the stability can be also seen in Refs.~\onlinecite{williams2002,navez2005} for a trap potential and uniform system, respectively.


\section{SUMMARY AND CONCLUSIONS}
In this paper, we have developed a coarse-grained finite-temperature theory for a Bose condensate in one-dimensional optical lattices, in addition to the confining harmonic trap potential. This theory consists of coarse-grained equations of motion for the condensate variables and noncondensate Green's functions, which include the effect of a dissipative term due to collisions between the condensate and the thermal cloud, as well as the noncondensate mean-field.

With use of the non-equilibrium field theory,  the 2PI effective action for the Bose condensate on the Schwinger-Keldysh closed-time path has been obtained.
Introducing an {\it ansatz} for the variational function in the effective action to perform a coarse-graining approximation, we have obtained a coarse-grained effective action, which includes the effects of the optical lattice potential effectively, in the presence of a thermal cloud of noncondensate atoms.
We have also derived a coarse-grained action in terms of hydrodynamic variables of the condensate.
Using the variational principle, we obtained coarse-grained equations of motion for the condensate variables, which can be used to describe the long wave-length dynamics on the length scale much longer than the lattice
constant $d$.

To illustrate our formalism, we used the generalized GP hydrodynamic equations to investigate the stability of
superfluidity in the current-carrying condensate. Following recent work~\cite{konabe2006_2,iigaya2006}, we calculated the damping rate of the collective oscillations.
We have found that the collisional damping rates change sign when the condensate velocity
exceeds the renormalized sound velocity, leading to the Landau instability consistent with the Landau criterion.
The results in this paper sheds light on the microscopic origin of the Landau instability. 

In the present paper, we concentrated on the effect of the optical lattice and ignored the trapping potential. One could use our formalism to analyze the experimental results by Florence group~\cite{burger2001,ferlaino2002} on the damping of the condensate in dipole oscillations due to the thermal cloud.

In order to describe the coupled non-equilibrium dynamics of both the condensate and noncondensate, one has to derive a kinetic equation for the noncondensate distribution function in the presence of an optical lattice.  
The generalized GP hydrodynamic equations for the condensate derived in the present paper and the kinetic equation for the noncondensate will be used as a sound
basis for investigating finite-temperature behaviors of the Bose condensate in optical lattices.

\section*{ACKNOWLEDGMENTS}
The authors thank Allan Griffin for useful suggestions and a critical reading  of the manuscript.
The authors also acknowledge valuable and helpful discussions with Ippei Danshita and Kiyohide Iigaya.
S. K. is supported by JSPS (Japan Society for the Promotion of Science) Research Fellowship for Young Scientists.


\end{document}